\documentclass{aastex63}

\received{May, 9, 2025}
\revised{June, 6, 2025}
\accepted{June, 28, 2025}
\submitjournal{ApJ}

\usepackage{graphicx}
\shorttitle{
On the origin of short-lived cocoon in 3C84: powered by TDEs?
}
\shortauthors{Kawakatu, Kino and Wada}

\begin{document}

\title{
On the origin of short-lived cocoon in 3C84: powered by tidal disruption events ?
}

\correspondingauthor{Nozomu Kawakatu}
\email{kawakatsu@kure-nct.ac.jp}

\author[0000-0003-2535-5513]{Nozomu Kawakatu}
\affiliation{Faculty of Natural Sciences, 
National Institute of Technology, Kure College, 2-2-11 
Agaminami, Kure, Hiroshima, 737-8506, Japan}
\affiliation{Kagoshima University, Graduate School of Science and Engineering, Kagoshima 890-0065, Japan}

\author[0000-0002-2709-7338]{Motoki Kino}
\affiliation{Kogakuin University of Technology \& Engineering, Academic Support Center, 2665-1 Nakano, 
Hachioji, Tokyo 192-0015, Japan
}
\affiliation
{National Astronomical Observatory of Japan, 2-21-1 Osawa, Mitaka, Tokyo 181-8588, Japan
}

\author[0000-0002-8779-8486]{Wada Keiichi}
\affiliation{Kagoshima University, Graduate School of Science and Engineering, Kagoshima 890-0065, Japan}
\affiliation{Ehime University, Research Center for Space and Cosmic Evolution, Matsuyama 790-8577, Japan}
\affiliation{Hokkaido University, Faculty of Science, Sapporo 060-0810, Japan}



\begin{abstract}
We evaluated the jet power and the density of ambient matter in 3C 84 by using the momentum balance along the jet axis and the transonic condition for the cocoons observed at two different scales (approximately 1 and 6 parsec scales).
For the inner cocoon, we precisely determined the ratio of jet power to ambient density $L_{\rm j}/n_{\rm a}$ to be $(0.3-0.7)\times 10^{43}\,{\rm erg}\,{\rm s}^{-1}\,{\rm cm}^3$. Similarly, for the outer cocoon, we found that this value is more than an order of magnitude larger at $(0.9-3.7)\times 10^{44}\,{\rm erg}\,{\rm s}^{-1}\,{\rm cm}^3$. This indicates that the outer cocoon is formed by a powerful jet that propagates through an ambient density of $20-300\,{\rm cm}^{-3}$ with a jet power of $10^{45-46.5}\,{\rm erg}\,{\rm s}^{-1}$. On the other hand, the inner cocoon is formed by a weaker jet with a power of $10^{43-44}\,{\rm erg}\,{\rm s}^{-1}$, propagating through a relatively low-density environment of $6-20\, {\rm cm}^{-3}$. 
These results suggest that: 1) with respect to the difference in $n_{\rm a}$, it appears to support the hypothesis that the inner cocoon, recently formed about 10 years ago, is expanding in the low-density cocoon created by the jet emitted about 25-50 years ago. 2) to achieve the short-lived and high $L_{\rm j}$ that generated the outer cocoon, a large mass accretion rate must be required over a short period to activate the jet. 
These may imply the extreme accretion event driven by the tidal disruption events (TDEs) of massive stars and/or the disk instability.
\end{abstract}

\section{Introduction} 
\label{sec:intro}
The formation of relativistic jets and their connection to the accretion disk 
remains a significant question in active galactic nuclei (AGN) physics 
\citep[e.g.,][]{Rees1984,Begelman84}. 
X-ray binaries, with their short dynamical timescales, provide a useful analogy for AGN jets. They show that the jet power may depend on the physical state of the accretion disk and the accretion rate \citep[e.g.,][]{Fender04, McClintock06}. The jet power in AGNs is correlated with narrow emission line luminosities, indicating a link between jet launching mechanisms and the accretion disk \citep[e.g.,][]{Baum89, Rawlings91}. However, interpreting this connection is challenging due to the long lifetimes of AGN jets, compared to rapid changes in accretion disk states \citep[e.g.,][]{ODea09}. Furthermore, estimating the power of AGN jets $L_{\rm j}$ is difficult because most emissions from AGN jets are from non-thermal electrons, obscuring signals from thermal and proton components. 

To solve this issue, jet power can be estimated by measuring cavity volumes and buoyant rise times for low-power Fanaroff-Riley I (FR I) radio galaxies \citep[e.g.,][]{Birzan04,Rafferty06,McNamara11, Russell13}. In contrast, high-power Fanaroff-Riley II (FR II) radio galaxies have cocoons with pressures exceeding the surrounding gas, allowing for jet power estimation through cocoon dynamics \citep[e.g.,][]{Begelman89,Kaiser97,Kino05, Ito08}. 
However, it is difficult to examine early jet formation stages, which is closely related to the physical state of the accretion disk, because these estimates are the averaged jet power over typical ages of FRIs and FRIIs, i.e., $10^{6-8}\,{\rm yrs}$. Therefore, estimating $L_{\rm j}$ for much younger compact radio galaxies is essential.

Compact symmetric objects (CSOs), whose sizes are from $10$ pc to $1$ kpc, are crucial for studying the early stage of jet formation, because CSOs with ages of $10^{2}$ to $10^{5}$ years are much younger than typical radio galaxies such as FRIs and FRIIs \citep[e.g., ][]{O'Dea98,Hardcastle20,OS21,Readhead24}. 
Observations show that CSOs represent a large fraction ($\sim 30\%$) in the flux-limited catalog of radio sources \cite[]{Fanti95}, which is much higher than the expected value ($ < 0.1 \%$) 
of their ages. This suggests that a large fractions of CSOs stop their jet activity at an early stage of their evolution $< 10^{5}\,{\rm yr}$ \citep[]{Reynolds97,Alexander00, Orienti00,Marecki03,Gugliucci05,Kunert-Bajraszewska10,Orienti10,Lister20,Nyland20,Kunert20,Woloska21}.
Recently, \citet{Kiehlmann24b} have shown that the edge-brightened high luminosity CSOs form a distinct subclass of jetted AGN that has a sharp upper cutoff in size at $\simeq 500\,{\rm pc}$ based on the complete sample \citep[][]{Kiehlmann24a}. 
Thus, many CSOs may be short-lived \citep[e.g.,][]{Readhead94,Readhead24}. Therefore, clarifying not only the jet power $L_{\rm j}$ but also the time variability of the jet power $L_{\rm j}(t)$ are essential to reveal the origins of CSOs. 

To obtain the total kinetic power $L_{\rm j}$ for CSOs, the estimation of ambient density around CSOs, $n_{\rm a}(r)$ plays a crucial role. However, it is challenging to estimate the ambient density within galaxies because the emission from diffuse ambient gas is very weak because of contamination from other bright sources.
According to the dynamical model of the expanding cocoon and the shell, the expansion velocity strongly depends on the ambient density \citep[e.g., ][]{Reynolds97,Reynolds02,Komissarov07, Nath10, Mocz11, Perucho11, Ito15}. Thus, if the expansion velocity is obtained from observations, it is possible to constrain the ambient density. 

\begin{figure}
\includegraphics[width=0.95\textwidth]{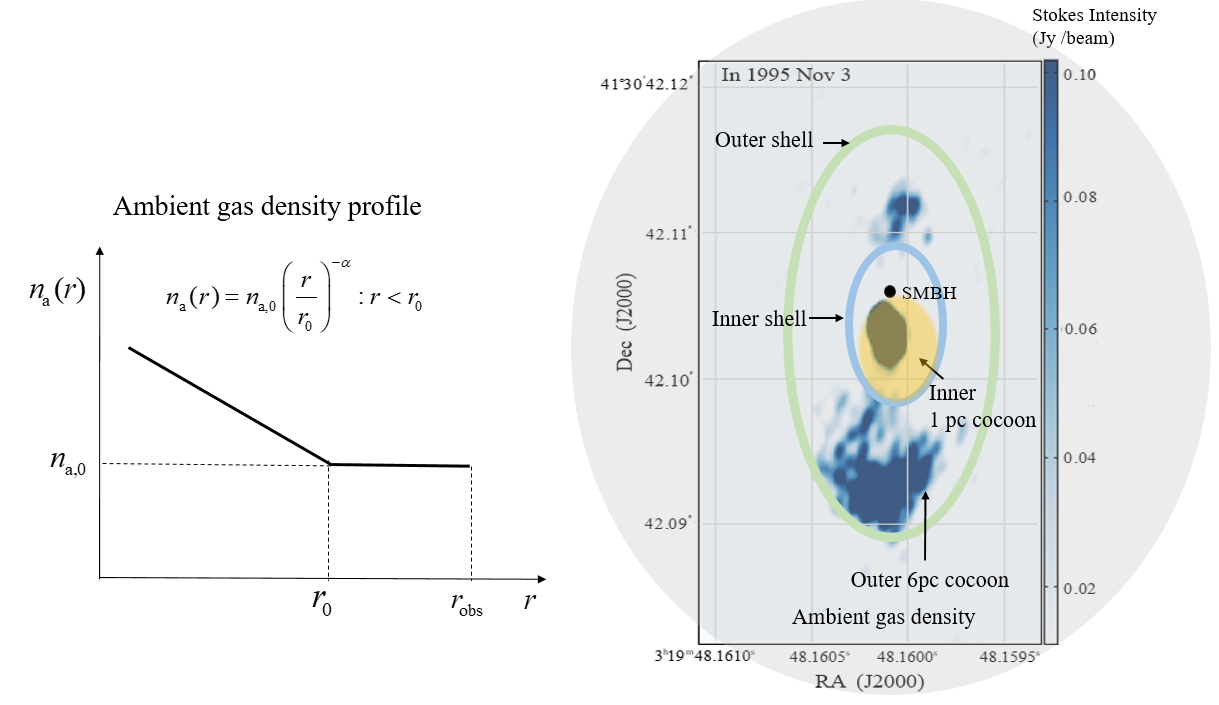}
\caption
{
Left: A sketch of the expected surrounding gas density profile. The vertical axis is the number density of ambient gas $n_{\rm a}(r)$. The horizontal axis is the distance from the SMBH, $r$. $n(r_{\rm a,0})=0.1\,{\rm cm}^{-3}$ is the density observed at $r=r_{0}=1\,{\rm kpc}$, $r_{0}$ is the critical radius of the gas density and $\alpha$ is the slope index of $n_{\rm a}(r)$ for $r < r_{0}$. 
Right: 
A cartoon of the multiple mini-cocoon/radio-lobe/shell system overlaid on the radio image of 3C 84.
The large gray circle represents the ambient gas surrounding the entire system.
The outer cocoon is depicted as white-colored bubbles. The current expanding velocity of outer cocoon/shell is subsonic (See Figure \ref{fig:MOJAVE_3epochs}  for details). As for the inner lobe, it is well known that the northern inner lobe is significantly affected by free-free absorption of ionized plasma near the central SMBH, making the northern inner lobe not clearly visible \citep[e.g.,][]{walker00,Wajima20}. Therefore, it is not depicted in this sketch. The small orange-colored bubble is the inner cocoon surrounding the parsec-scale newborn jet observed using 5GHz {\it RadioAstron} space-VLBI observations \citep{Savolainen23}. The expanding velocity of inner cocoon/shell is still supersonic. The blue and green rings represent the 
inner and outer shells, respectively.
\label{fig:1}
}
\end{figure}

The nearby bright radio galaxy 3C 84 ($z = 0.0176$), located at the center of the Perseus cluster and harboring a supermassive black hole (SMBH) with a mass of $M_{\rm BH} = (0.8-2)\times 10^{9}M_{\odot} $
\citep[]{Scharwachter13,Giovannini18}, provides an ideal environment for studying the physics of energy transport by radio lobes at parsec scales. Since a pair of jets has been observed on scales ranging from sub-pc to 10 kpc scale \citep[e.g., ][]{Morganti23}, we recognize that 3C 84 is an extreme case which exhibits intermittent jet ejections.
In the central region of 3C 84, single-dish observations of 3C 84 showed that the outburst started in 1959 and found two types of variations; one is fast outbursts happen at radio frequencies 5 years after the corresponding optical flares. The rapidly varying lasts from several months in the optical and a few years at radio band. The other is the slowly varying component which has a timescale of about 30 years. 
This slow variation leads to the growth and decay of the radio emitting jets \citep[e.g.,][]{Nesterov95}. 
By using a six-station VLBI network, \citet{Readhead83} revealed compact radio structure, which may be related to the radio outburst in 1959.
Recently, much more compact radio sources were discovered within the outer 6 pc-scale radio cocoon (hereafter we call it the outer cocoon) observed by VLBI Space Observatory Program (VSOP) observations\citep[]{Asada06}. 
The C3 component in 3C 84 is identified as a newborn radio lobes started in 2005 and it propagates southward \citep[e.g.,][]{Nagai10}. 

These radio morphologies in the central 10 pc scale region are quite similar to that of CSOs. Thus, it is useful to understand the origin of CSOs because the compact ($< 10\,{\rm pc}$) cocoons may be the early phase of more extended CSOs whose seize is from 10 pc to kpc.  
Recently, using 5GHz {\it RadioAstron} space-VLBI observations, \citet{Savolainen23} reported the first discovery of a cocoon-like, low-intensity, emission structure surrounding the parsec-scale restarted jet of the 3C 84 (hereafter we call it the inner cocoon). 
These findings indicate that we can investigate whether there are variations in jet power over a short period for the inner and outer cocoons in 3C 84, $L_{\rm j}(t)$. 
However, this has been challenging due to significant uncertainties in the surrounding ambient gas density within $<\, 10{\rm pc}$ in order to examine the jet power on a parsec scale. In this paper, we will assess the density of ambient gas using the expansion velocity of the cocoon and the shell, as seen in Figure \ref{fig:1}. 

The goal of this paper is to estimate the kinetic jet power and ambient density around parsec-scale multiple jets in 3C 84, based on the conditions of supersonic and subsonic expansion of cocoons.
In Section \S \ref{sec:model}, the method, problem setting, and dynamical model in the AGN cocoon and shell are briefly described. In Section \S \ref{sec:results}, we present the results. We provide a discussion in \S \ref{sec:discussions}. The summary is given in Section \S \ref{sec:summary}.
With the cosmology parameters of $\Omega_{\rm m} = 0.27$, $\Omega_{\Lambda} = 0.73$, and $H_{0} = 71 {\rm km}\, {\rm s}^{-1}\, {\rm Mpc}^{-1}$ 
\citep[]{Komatsu09} the angular scale of 1 mas corresponds to a linear distance of 0.35 pc for 3C 84.

\section{Model of cocoon/shell expansion in 3C 84}\label{sec:model}
\subsection{Problem setting and basic idea of the method}
Details of the dynamical model of over-pressured cocoon’s expansion \citep[e.g.,][]{Begelman89,Kino05} and shell \citep[]{Ito15} have been established. In this paper, we apply these well-established models for simplicity. In models, the physical quantities of the cocoon and shell are expressed as functions of $L_{\rm j}$ and the age of the cocoon $t_{\rm age}$. However, here we can treat the physical quantities of cocoon and shell as given by $L_{\rm j}$ because $t_{\rm age}$  of mini-cocoon is well constrained by observations.

\begin{deluxetable*}{cccccc}[ht]
\tablecaption{Physical quantities in 3C84 \label{table:1}}
\tabletypesize{\small}
\tablehead{
  \colhead{Quantities} &
  \colhead{Symbol}
&    \colhead{Note} 
  }
\startdata
Jet kinetic power & $L_{\rm j}$ & model parameter \\
Distance from the SMBH to the cocoon head & $R_{\rm h}$ & 1.4 - 4 pc for inner cocoon \\
& & 5.6 - 15.7 pc for outer cocoon \\
Velocity of cocoon head & $v_{\rm h}(r)$ & observable parameter \\
Velocity of cocoon head at $r=r_{0}$ & $v_{\rm h, 0}$ & observable parameter \\
Cross-sectional area of cocoon/lobes & $A_{\rm h}(r)$ & observable parameter \\
Cross-sectional area of cocoon/lobes at $r=r_{0}$ & $A_{\rm h, 0}$ & observable parameter \\
Shell/cocoon radius & $R(t)$ & model parameter \\
Expanding shell velocity & $\dot{R}(t)$ & model parameter \\ 
Growth rate of the cocoon head & $\beta$  &  
1 (parabolic case) , 2 (conical case)\\\hline 
Age of cocoons & $t_{\rm age}$ & observable parameter \\
Duration of jet injection & $t_{\rm j}$ &  free parameter \\
Transonic timescale & $t_{\rm trans}$ & Eq. (7) \\\hline
Number density of ambient gas  & $n_{\rm a}(r)$ & model parameter \\
Number density of ambient gas at $r=r_{0}$ & $n_{\rm a,0}$  &  0.1 cm$^{-3}$ at $r=r_{0}$\\
Critical radius of ambient gas density & $r_{0}$ 
& $10\,{\rm pc}$, $100\,{\rm pc}$ and $1\,{\rm kpc}$ \\  
Power-law index of ambient gas density & $\alpha$ & 0.5--2 \\
\enddata
\end{deluxetable*}

In Figure \ref{fig:1}, we present a cartoon of the multiple mini-cocoon/shell system observed in 3C 84. The kinetic energy of the jets is dissipated through the termination shock at the hot spots and deposited into the cocoon and the shell with its width. The cocoon is inflated by its internal energy. The cocoon drives the forward shock propagating in the ambient gas, and the forward-shock region forms the shell. 
The model parameters are the jet power, $L_{\rm j}$ and the number density of the spherically symmetric ambient gas, $n_{\rm a}(r)$ where $r$ is the distance from the SMBH. Here, $L_{\rm j}$ takes into account the possibility of a change between the formation of the two cocoons, but $L_{\rm j}$ is assumed to be constant during each cocoon formation. 
Table \ref{table:1} summarizes the physical quantities used in our model. Among them, the obtained quantities are the velocity of the cocoon head $v_{\rm h}(R_{\rm h})$, the cross-sectional area of the cocoon $A_{\rm h}(R_{\rm h})$, and the age of the cocoon $t_{\rm age}$, where $R_{\rm h}$ is the distance from the SMBH to the cocoon head. 
The free parameters are the critical radius $r_{0}$ and the power law index $\alpha$, 
which determine the profile of the ambient gas density (see Table \ref{table:1}). 
Thus, we can constrain $L_{\rm j}$ and $n_{\rm a}(R_{\rm h})$, since the basic equations are two: the momentum balance along the jet axis and the subsonic condition of the expanding shell. 

The method is divided into four steps as follows:
(i) In Section 2.2, we analyzed three epochs of VLBA archival data of 3C 84 at 15 GHz obtained in 1995, 2015 and 2022 to understand the basic characteristics of both inner and outer cocoons.
(ii) We specify the properties of the surrounding gas in Section 2.3, assuming that the ambient gas density is closely related to the evolution of the shell/cocoons. 
(iii) The ratio of the total jet power to the ambient density, $L_{\rm j}/n_{\rm a}(R_{\rm h})$ is determined by the momentum balance along the jet axis in Section 2.4. Using the observed properties of cocoons, namely, the advance speed of the cocoon head, $v_{\rm h}$ and the cross-sectional area of the cocoon, $A_{\rm h}$ and the age of the cocoon $t_{\rm age}$, we constrain $L_{\rm j}/n_{\rm a}(R_{\rm h})$ for each cocoon.  
(iv) We constrain the ambient gas density under the condition that the velocity of the expanding shell/cocoon is the sound velocity of the surrounding gas in Section 2.5. Recent observations suggest that the inner cocoon expands supersonically, while a pair of fading outer radio lobes expands subsonically, as shown in Figure \ref{fig:MOJAVE_3epochs}.
Finally, the jet power $L_{\rm j}$ for the inner and outer cocoons is used to estimate how $L_{\rm j}(t)$ changes in time.

\subsection{Inner and outer cocoons of 3C 84 images at 15~GHz}
\begin{figure}
\includegraphics[width=0.95\textwidth]{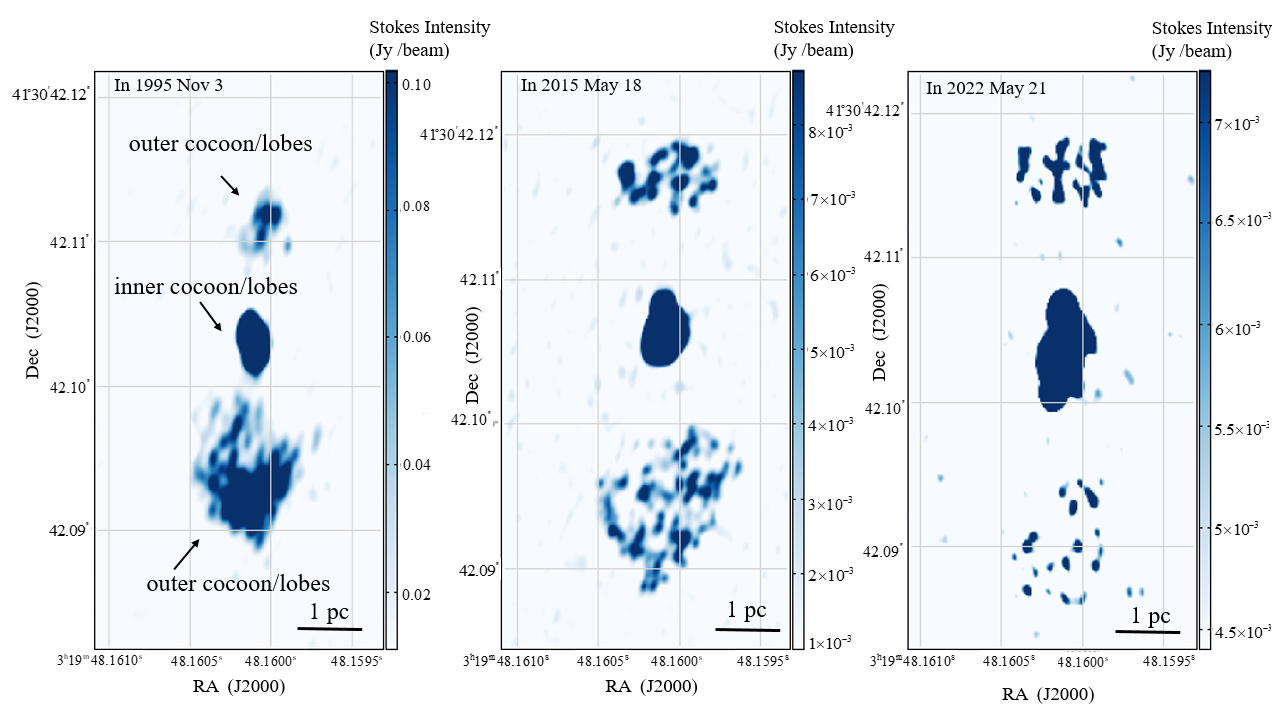}
\caption
{
Comparison of the three epochs of the actual 
3C 84 images at 15~GHz taken from the MOJAVE project (https://www.cv.nrao.edu/MOJAVE/)
in 1995 Nov 3, 2015 May 18, and 2022 May 21, 
with the project IDs, BA12, BL193AS, BL286AK, respectively. 
The image rms (1 $\sigma$) are
11.12~mJy/beam (BA12), 
1.04~mJy/beam (BL193AS),
4.44~mJy/beam (BL286AK), respectively.
Typical spatial resolution is the dimensions of 1.1 mas $\times$0.5 mas \citep[]{Lister18}, 
where the angular scale of 1 mas corresponds to a linear distance of 0.35 pc.
Corresponding to Figure~\ref{fig:1},
the expanding central compact radio lobe with the size of 1 pc order (inner lobes/cocoon) and a pair of fading outer radio lobes with the size of 6 pc order (outer lobes/cocoon) are seen in these images.}
\label{fig:MOJAVE_3epochs}
\end{figure}

To understand the basic characteristics of the cocoons, we displayed three epochs of relatively good quality VLBA archival data of 3C 84 at 15 GHz obtained in 1995, 2015,
and 2022 (project IDs are BR12, BL193AS, and
BL286AK, respectively). These three were adopted from
already analyzed MOJAVE images (http://www.physics.purdue.edu/MOJAVE/).
In Figure \ref{fig:MOJAVE_3epochs}, we show the obtained intensity map of the inner and outer radio cocoon/lobes. 
The image root mean square (rms) (1 $\sigma$) values are
11.12 mJy/beam (BA12), 1.04 mJy/beam (BL193AS),
4.44 mJy/beam (BL286AK), respectively. 
The typical spatial resolution is the dimensions of 1.1 mas $\times$0.5 mas \citep[]{Lister18}. 
We find that the total flux of the northern and southern lobes has decreased tenfold over 20 years. In addition, we observe that the outer radiolobe morphology started to collapse after the epoch in 2015, and the expansion speed of the cocoon head should be subsonic because there is no evidence of lobe expansion from 2015 to 2021 \citep{kino17}. On the other hand, recent observations indicate that the expanding velocity of the inner cocoon is super-sonic \citep[e.g.,][]{Nagai10}. As shown in Figure~\ref{fig:1}, the expanding central compact radio lobe with a size of approximately 1 pc (inner cocoon) and a pair of fading outer radio lobes with a size of about 6 pc (outer cocoon) are seen in these images. 

\subsection{Density profile of ambient gas}
To study the evolution of mini-cocoon/shell, we need to specify the properties of the surrounding gas. Using the deep Chandra observation \citep[]{Fabian06}, \citet{Taylor06} estimated that the number density at $r_{\rm obs}=1$ kpc
is $\simeq 0.1\,{\rm  cm}^{-3}$. 
However, there is no information on the gas for $r < 1$ kpc because the hot gas is strongly affected by the X-ray cavities created by past AGN activities and strong AGN feedback in the vicinity of the SMBHs. 
As seen in Figure \ref{fig:1}, we postulate the density profile of ambient gas, $n_{\rm a}(r)$ as follows; 
\begin{eqnarray}\label{eq.density}
n_{\rm a}(r) &=&
\left \{ 
 \begin{array}{l}
 n_{\rm a,0}\left(\frac{r}{r_{0}}\right)^{-\alpha}; \,\,\,r < r_{0}\\
 n_{\rm a,0}\, ; \,\,\, r_{0} \le  r  < r_{\rm obs}, 
 \end{array}
\right.
\end{eqnarray}
where $\alpha$ is the slope index of $n_{\rm a}(r)$. 
Here, $n_{\rm a,0}=0.1\,{\rm cm}^{-3}$ is the ambient gas density at $r=r_{0}$, and $r_{0}$ is the critical radius of the ambient gas density. Concerning the critical radius $r_{0}$, we adopt three cases; the case (i) is $r_{0}=10\,{\rm pc}$, case (ii) is $r_{0}=100\,{\rm pc}$, and case (iii) is $r_{0}=10^3{\rm pc}$. 
In this paper, we focus on the diffuse hot gas that mainly interacts with the expanding mini-cocoon. 
The region along the jet axis is considered to be essentially free of gas clouds because there is no centrifugal force acting on gas clouds \citep[e.g., ][]{Ramos17}. Thus, the clumpy gas clouds do not affect on the global jet propagation significantly, although 
a local collision between the jet and a compact dense clouds at the central one-parsec region has been reported \citep[]{Kino21,Kam24}. 
We also assume the hot gas temperature around the galactic center ($r < 1\,{\rm kpc}$) is constant with 
$T_{\rm a}=3\,{\rm keV}$ which is the gas temperature at $r\sim 1\,{\rm kpc}$ \citep[]{Fabian06}. 
The dependence of $T_{\rm a}$ will be discussed in Section 3.1.2. 

For cases (i) and (ii), we propose a flat density profile for $r_{0} < r < r_{\rm obs}$, which may reflect a scenario in which the thermal instability of the gas is effective and most of the hot gas becomes cold gas
\citep[e.g.,][]{Barai12, Gaspari12, McCourt12, Sharma12, Guo14, Meece15}. For the slope index $\alpha$, we examine a wide range of parameter space, i.e., $\alpha=0.5-2$. For the classical Bondi accretion, $\alpha$ is expected to be 1.5 from the mass conservation. On the other hand, in the context of advection-dominated inflow-outflow solution (i.e.,  the radiatively inefficient accretion flows (RIAF)-type hot accretion flows with substantial winds or mass loss), $0.5 \leq \alpha \leq 1.5$ was suggested by \cite{Blandford99}. 
Moreover, numerical simulations of a gravitational collapse in the presence of the gas cooling have shown that the slope is $0 < \alpha < 2$ \citep[e.g.,][]{Choi13,choi15}. 
For reference, we define characteristic radius such as the Bondi radius, $r_{\rm B}$ and the radius of the sphere of gravitational influence (SGI), $r_{\rm SGI}$. Suppose that the central stellar velocity dispersion is $\sigma = 250\,{\rm km/s}$ \citep[]{Bettoni03} and the mass of the central black hole is $M_{\odot} = 8\times 10^{8} M_{\odot}$ \citep[]{Scharwachter13}, $r_{\rm B}$ and $r_{\rm SGI}$ are $r_{\rm B}=2GM_{\rm BH}/c_{\rm s}^{2}=12\,{\rm pc}$ and $r_{\rm SGI}=2GM_{\rm BH}/\sigma^{2}=60\,{\rm pc}$ where $c_{\rm s}$ is the sound velocity of the ambient gas with $T_{\rm a}$.

\subsection{Evolution of the cocoon}
The over-pressured cocoon model was initially proposed by \citet{Begelman89}, where the dissipated energy of the jet bulk motion is the origin of the total pressure of the cocoon \citep[e.g.,][]{Kaiser97, Kino05, Fujita16}. 
A supersonic cocoon is expected to be over-pressured against ambient pressure with a significant sideways expansion.
Thus, in general, we need to solve three basic equations: the equation of motion along the jet axis, sideways expansion, and energy conservation in the cocoon. However, thanks to the well-constrained quantities of the mini-cocoon by observations (see Table \ref{table:1}), it is sufficient to consider only the evolution of the cross-sectional area of the cocoon $A_{\rm h}$ and that of the expanding speed of the cocoon $v_{\rm h}$. Since $v_{\rm h}$  is about $10\%$ of light speed $c$ \citep[e.g.,][]{KKH08}, the relativistic effect on the cocoon evolution is negligible.
The momentum balance along the jet is given by :  
\begin{eqnarray}\label{mom}
\frac{L_{\rm j}}{v_{\rm j}} =m_{\rm p}n_{\rm a}(r){v_{\rm h}(r)}^{2}A_{\rm h}(r),
\end{eqnarray}
where $L_{\rm j}$, $v_{\rm j}$, $m_{\rm p}$, $v_{\rm h}$ and $A_{\rm h}$ 
are the kinetic power of jets, the velocity of jet, the proton mass, the advance velocity of cocoon, and the cross-sectional area of cocoon head, respectively. 
We assume that the velocity of jets are relativistic speed, i.e., $v_{\rm j}=c$  because the bulk Lorentz factor $\Gamma=(1-(v/c)^{2})^{-0.5}$ is larger than 3 \citep[e.g.,][]{Giovannini18}.
The growth rate of $A_{\rm h}$ is assumed to be merely a power law as 

\begin{eqnarray}
A_{\rm h}(r)=A_{\rm h, 0}
\left(
\frac{r}{r_{\rm 0}}
\right)
^{\beta},
\end{eqnarray}
where $\beta$ is the power law index of $A_{\rm h}(r)$. We here assume two cases: 
$\beta=1$ is the parabolic case and $\beta=2$ is the conical case. 
Thus, the ratio of jet power to ambient density is given by 
$L_{\rm j}/n_{\rm a}(r) = m_{\rm p}{v_{\rm h}(r)}^{2}A_{\rm h}(r)c.$ 
Note that the physical quantities $A_{\rm h}$ and $v_{\rm h}$ are observable in the right hand of above equation (Eq.(\ref{mom})). 
In addition, the velocity of the cocoon head is expressed as 
\begin{eqnarray}
v_{\rm h}(r)=v_{\rm h, 0}
\left(
\frac{r}{r_{0}}\right)^{\frac{\alpha-\beta}{2}}.
\end{eqnarray}

From Eqs. (2) ,  (3) and (4), we obtain the time $t$ as a function of $r$ as   
\begin{eqnarray}\label{age}
t(r)=\frac{2r}{2-\alpha+\beta}
\left(
\frac{L_{\rm j}}{m_{\rm p}n_{\rm a}(r)A_{\rm h}(r)c}
\right)^{-1/2}, 
\end{eqnarray}
which is consistent with the results \citep[]{Fujita16, Fujita17} for $\beta=2$.
We should note that the age of the cocoon $t_{\rm age}$ is the time at $r=R_{\rm h}$.
 
\subsection{Evolution of shell}
To treat the dying phase of the outer cocoon as seen in Figure \ref{fig:MOJAVE_3epochs}, we consider the evolution of the expanding shell. After stopping jet energy injection, that is, $L_{\rm j}=0$ for $t > t_{\rm j}$, the cocoon will rapidly lose its energy due to adiabatic expansion and transfer most of its energy to the shell within a dynamical timescale, where $t_{\rm j}$ denotes the duration of jet injection \citep[]{Ito15}. 
For the phase after the jet has been turned off ($t > t_{\rm j}$), the expansion of the bow shock (shell) is expected to be similar to the Sedov-Taylor solutions. The radius of the shell $R(t)$ is given by 
\begin{eqnarray}
R(t)=R(t_{\rm j})\left(\frac{t}{t_{\rm j}}\right)^{\frac{2}{5-\alpha}}. 
\end{eqnarray}
Here $R(t_{\rm j})=C(L_{\rm j}/(m_{\rm p}n_{\rm a,obs}))^{1/(5-\alpha)}r_{\rm 0}^{\alpha/(\alpha-5)}$, where $C$ is the numerical coefficient and the explicit form of $C=[(3-\alpha)(5-\alpha)^{3}(\gamma_{\rm c}-1)/
{4\pi(2\alpha^{2}+\alpha-18\gamma_{\rm c}\alpha+63\gamma_{\rm c}-28}]^{1/(5-\alpha)}
$ where $\gamma_{\rm c}=4/3$ since plasma within the cocoon is expected to be relativistic. 
On the other hand, the transonic condition of the expanding shell velocity 
$\dot{R}(t=t_{\rm trans})$ is 
\begin{eqnarray}\label{R}
\dot{R}(t=t_{\rm trans})=c_{\rm s},
\end{eqnarray}
where 
\begin{eqnarray}\label{t_tr}
t_{\rm trans}=\left(\frac{C}{c_{\rm s}}\right)^{\frac{5-\alpha}{3-\alpha}}
\left(\frac{L_{\rm j}}{m_{\rm p}n_{\rm a,0}r_{0}^{\alpha}}\right)^{\frac{1}{3-\alpha}}t_{\rm j}^{\frac{1}{3-\alpha}}.  
\end{eqnarray}
The sound velocity of the ambient gas is given by
$c_{\rm s}=(k_{B}T_{\rm a}/\mu m_{\rm p})^{1/2}$ where $k_{\rm B}$ and $\mu=0.6$ are the Boltzmann constant and the mean molecular weight of the ambient gas. Here we assume that $c_{\rm s}$ is constant.  
Combining Eqs.(\ref{R}) and (\ref{t_tr}), the above transonic condition  
can be rewritten as
\begin{eqnarray}
\left(\frac{L_{\rm j}}{n_{\rm a,0}}\right)_{\rm trans}
=C^{\alpha-5}c_{\rm s}^{2}
m_{\rm p}R_{\rm h}^{3}t_{\rm j}^{-1}
\left(\frac{r_{0}}{R_{\rm h}}\right)^{\alpha}, {\rm where} 
\,R(t=t_{\rm trans}) = R_{\rm h}. 
\end{eqnarray}

The velocity of the shell is expressed as $(E_{\rm j}/M_{\rm s})^{1/2}$ 
where the total energy $E_{\rm j}=L_{\rm j}t_{\rm j}$ and the swept mass of the shell $M_{\rm s}$. Thus, the dependence of $(L_{\rm j}/n_{\rm a, 0})_{\rm trans}$ on $t_{\rm j}$ and $R_{\rm h}$ is easily understood. 
As $r_{0}$ and $\alpha$ become larger, $n_{\rm a}(R_{\rm h})$ increases. Since larger $n_{\rm a}(R_{\rm h})$ cause a strong deceleration of the shock velocity, higher $(L_{\rm j}/n_{\rm a, 0})_{\rm trans}$ is required for given $R_{\rm h}$.  

\section{Results}\label{sec:results}
We first constrain the ratio of jet power to ambient gas density $L_{\rm j}/n_{\rm a}(R_{\rm h})$ using the momentum balance along the jet axis. Applying the transonic condition, we constrain the slope index $\alpha$ and the critical radius $r_{0}$ of the ambient gas density. 
Next, we evaluate the ambient gas density profile $n_{\rm a}(r)$ for both inner and outer cocoons. 
Finally, we estimate the jet power, $L_{\rm j}$, associated with the inner and outer cocoons and discuss how the jet power in 3C 84 changes in the last 50 years.

\begin{figure}
\includegraphics[width=0.95\textwidth]{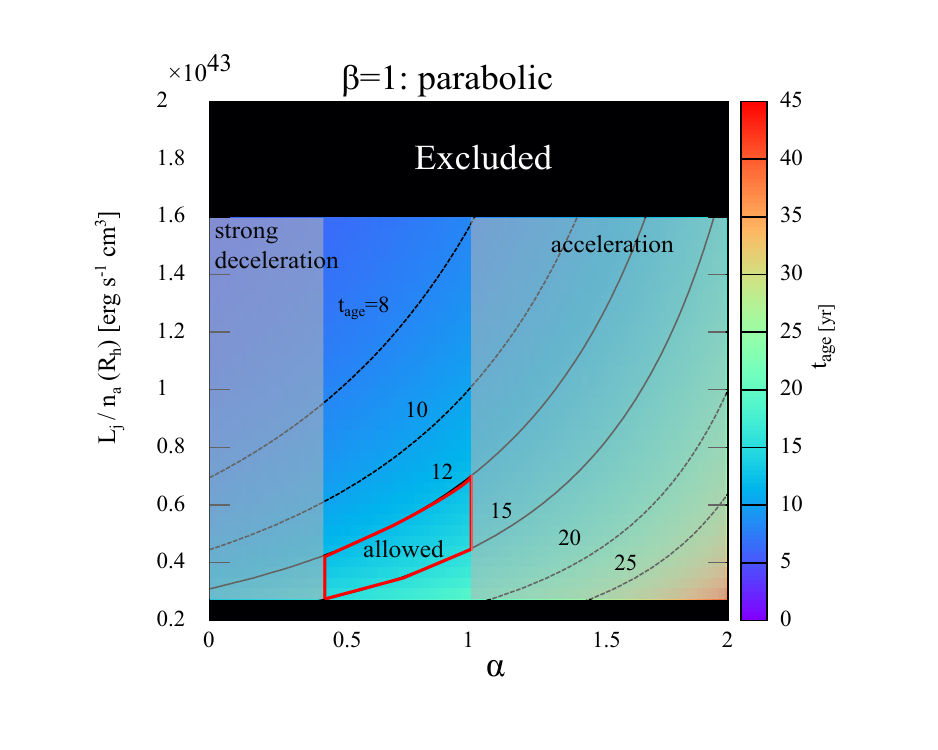}
\caption
{
Contour plots of inner cocoon ages, $t_{\rm age}$ as functions of the ratio of jets and ambient density at $R_{\rm h}$, $L_{\rm j}/n_{\rm a}(R_{\rm h})$ and the slope index of ambient density, $\alpha$ for the growth rate of the cocoon head with $\beta=1$ (parabolic shape). The black lines represent different ages with $t_{\rm age}=8, 10, 12, 15, 20 {\rm and}\, 25\,{\rm yr}$, respectively. 
The red outline represents the allowed range of $L_{\rm j}/n_{\rm a}(R_{\rm h})$ 
and $\alpha$ of the inner cocoon in 3C 84. 
The black filled area denotes the excluded range of $L_{\rm j}/n_{\rm a}(R_{\rm h})$ by the momentum balance along the jet axis. The light gray areas are excluded from the range of $\alpha$ because $v_{\rm h}$ is almost constant.
} 
\label{fig:3}
\end{figure} 
\begin{figure}
\includegraphics[width=0.95\textwidth]{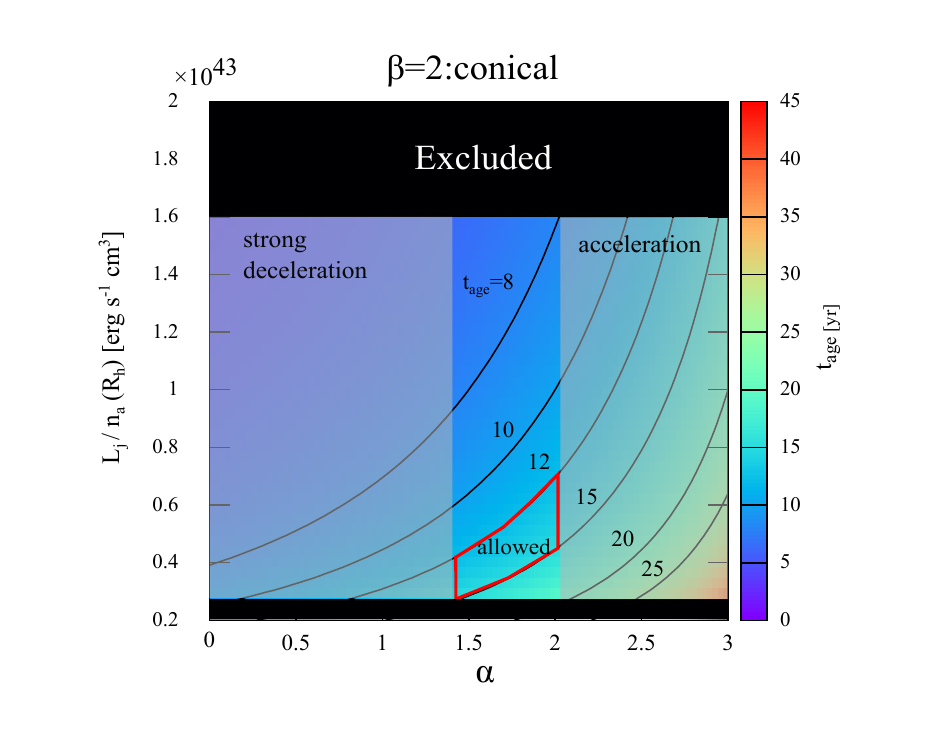}
\caption
{
Same as Figure \ref{fig:3}, but  for the growth rate of the inner 
cocoon head with $\beta=2$ (conical shape).
}
\label{fig:4}
\end{figure}

\subsection{Constraint on $L_{\rm j}/n_{\rm a}(R_{\rm h})$ and $n_{\rm a}(R_{\rm h})$ }
\subsubsection{Allowed $L_{\rm j}/n_{\rm a}$}
Figure \ref{fig:3} shows the ratio of $L_{\rm j}/n_{\rm a}(R_{\rm h})$ as a function of the ambient gas density profile, $\alpha$ and the age, $t_{\rm age}$ for a given jet head growth rate ($\beta=1$: parabolic shape). The black lines represent different ages of the inner cocoon with $t_{\rm age}=8,\,\,10,\,\,12,\,\,15,\,\, 20,\,{\rm and}\,25\,{\rm yr}$, respectively (see Eq. (5)). 
Only by the momentum balance along the jet (Eq.(\ref{mom})), the allowed region is obtained as $L_{\rm j}/n_{\rm a}(R_{\rm h})=(0.3-1.6)\times 10^{43}{\rm ergs s^{-1} cm^{-3}}$ using the observed values with  $v_{\rm h}=0.23-0.55\,c$ and $A_{\rm h}=0.13-0.39\,{\rm pc}^{2}$ at $R_{\rm h}=1.4-4\,{\rm pc}$ \citep[]{Savolainen23} for the wide range of the jet inclination angle $\theta =18^{\circ}-65^{\circ}$
\citep[]{Lister09,Nagai10,Suzuki12, Fujita17} 
Next, we further constrain the ratio of $L_{\rm j}/n_{\rm a}(R_{\rm h})$ and the slope index $\alpha$ by incorporating the allowed range of the observed cocoon age $t_{\rm age}$ and the advance velocity of the cocoon $v_{\rm h}(t)$. 
We estimate the age of the inner cocoon $t_{\rm age}$ in September 2013, when the 5 GHz image was taken by {\it Radioastron}. According to \citet{Suzuki12}, the new-born hot spot "C3" was first identified in November 2003 and the initial velocity is $0.09\pm 0.04\,{\rm mas/yr}$. In addition, the physical distance from the SMBH is 0.15 mas, considering the effect of the core shift \citep[]{Paraschos21}.
Combining these values, the age of the inner cocoon can be very constrained to $t_{\rm age}=12-15\,{\rm yr}$. 

For $v_{\rm h}(t)$, solutions exhibiting large deviations from constant velocity are safely ruled out from the VLBI observations \citep[]{Nagai10,Suzuki12}. Thus, we impose the condition that the expansion speeds $v_{\rm h}$ at 0.1 pc must be smaller than $c$ for $v_{\rm h}(R_{\rm h})=0.5\,c$. As a result, we find that a smaller slope index, $\alpha=0-0.4$, corresponds to a rapid deceleration in $v_{\rm h}$, while a larger $\alpha=1-2$ denotes an acceleration in $v_{\rm h}$ (see Eq. (4)). Thus, we can constrain the allowed range of the slope index $\alpha$ to $\alpha=0.4-1$. 

Combining the momentum balance along the jet axis, the observed range of $t_{\rm age}$ and the allowed range of $\alpha$, we tightly constrain $L_{\rm j}/n_{\rm a}(R_{\rm h})$. 
The allowed range of $L_{\rm j}/n_{\rm a}(R_{\rm h})$ is obtained as 
\begin{eqnarray}
\left(\frac{L_{\rm j}}{n_{\rm a}(R_{\rm h})}\right)_{\rm inner}
=(0.3-0.7)\times 10^{43}\,{\rm ergs\, s^{-1}\, cm^{-3}}, 
\end{eqnarray} 
which is highlighted as the red area in Figure \ref{fig:3}. 
Figure \ref{fig:4} shows the allowed range of $L_{\rm j}/n_{\rm a}(R_{\rm h})$ with the growth rates of the jet head ($\beta=2$:conical shape). 
Similarly, the allowed range of $\alpha$ is $\alpha=1.4-2$. Compared to Figure \ref{fig:4}, we confirm that the allowed rage of $L_{\rm j}/n_{\rm a}(R_{\rm h})$ shows little dependence on the growth rate of $A_{\rm h}$, i.e., $\beta$. 

\citet{Asada06} conducted VLBI Space Observatory Program (VSOP) observations of 3C 84, revealing details of the structure of radio lobes. Given that we can obtain key physical quantities such as $A_{\rm h}(R_{\rm h})=2-5.9\,{\rm pc}^{2}$ and $v_{\rm h}(R_{\rm h})=0.2-0.4\,c$ where $R_{\rm h}=5.6-15.7\,{\rm pc}$, 
the ratio of $L_{\rm j}/n_{\rm a}$ can be determined by the momentum balance along the jet axis (see Eq.(\ref{mom})) as

\begin{eqnarray}
\left(\frac{L_{\rm j}}{n_{\rm a}(R_{\rm h})}\right)_{\rm outer}
=(0.9-3.7)\times 10^{44}\,{\rm ergs\, s^{-1}\, cm^{-3}}, 
\end{eqnarray}
which is an order of magnitude larger than that of the inner cocoon and is consistent with previous studies \citep[]{Fujita16}. 

\begin{figure}
\includegraphics[width=0.95\textwidth]{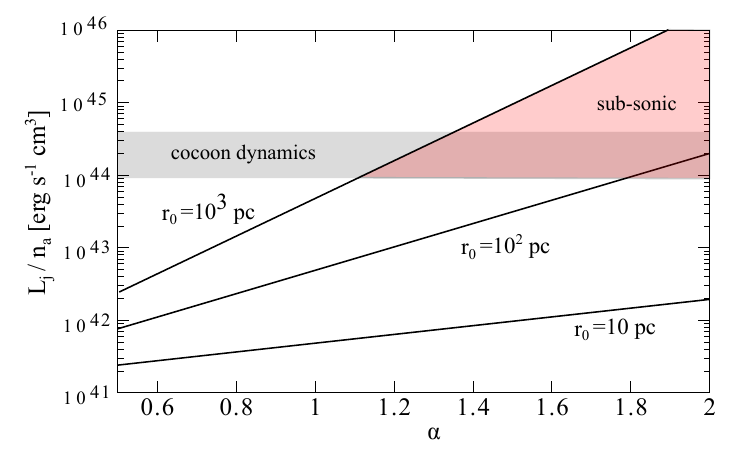}
\caption
{
Condition of the subsonic expansion of the outer cocoon with $t_{\rm j}=50\,{\rm yr}$ and $R_{\rm h}=6\,{\rm pc}$. 
The black lines represent the ratio of jet power and ambient density for the transonic condition (Eq. (9)) against $\alpha$ for 
the different critical radius, $r_{\rm 0}=10\,{\rm pc}$, $100\,{\rm pc}$ and $10^{3}\,{\rm pc}$, respectively.
The red area shows the allowed area, which is satisfied with the the condition  of the subsonic expansion. 
The grey area denotes the allowed $L_{\rm j}/n_{\rm a}(R_{\rm h})$ estimated by the momentum balance along the jet axis.
}
\label{fig:5}
\end{figure} 

\subsubsection{Allowed $\alpha$}
Figure \ref{fig:5} shows the ratio of jet power and ambient density, $L_{\rm j}/n_{\rm a}(R_{\rm h})$,  as a function of the slope index of ambient density, $\alpha$ for the outer cocoon. 
The black lines show the transonic condition (Eq. (9)) for different critical radii $r_{0}=10\,{\rm pc}$, $100\,{\rm pc}$ and $10^{3}\,{\rm pc}$ 
with $t_{\rm j}=t_{\rm age}=50\,{\rm yr}$ and $R_{\rm h}=6\,{\rm pc}$. 
It is hard to accurately estimate the duration of the jet
injection ($t_{\rm j}$) or, equivalently, the time when the jet stopped because it is not a direct observable. However, we shortly discuss the lower value of $t_{\rm j}$. The outer fading cocoon clearly indicates that $t_{\rm j}$ is shorter than $t_{\rm age}$ of the outer cocoon, where the range of $t_{\rm age}$ is $\simeq 45-60\,{\rm yr}$ \citep[]{kino17}. For the lower value of $t_{\rm j}$, since the radio flux of the outer cocoon started to decline 25 -30 years later \citep[]{Nesterov95}, $t_{\rm j}=25-30\,{\rm yr}$ could be expected. On the other hand, a theoretical point of view, if $t_{\rm j}\leq 0.5t_{\rm age}$, the radio flux of the cocoon decrease by an order of magnitude because of the Synchrotron cooling \citep[]{Ito15}. Thus, the lower value of $t_{\rm j}=25\,{\rm yr}$ would be reasonable. Even if a shorter duration, such as $t_{\rm j} =25\,{\rm yr}$, is adopted, our results do not change significantly because of $(L_{\rm j}/n_{\rm a})_{\rm trans}\propto t_{\rm j}^{-1}$(see Eq. (9)).
If the observed value $L_{\rm j}/n_{\rm a}$ is located above the line of $t_{\rm age} = 50\,{\rm  yr}$ for given $r_{\rm 0}$ and $\alpha$, it indicates that the expansion is subsonic (see Figure \ref{fig:MOJAVE_3epochs}).
The gray shaded area represents the allowed range of $L_{\rm j}/n_{\rm a}$ derived from the cocoon dynamics, and the red-hatched region marks the areas satisfying the subsonic expansion conditions.
As a result, we find that the slope index $\alpha=1.1-2$ and 
$\alpha=1.8-2$ are satisfied with the sub-sonic conditions for $r_{0}=10^{3}{\rm pc},\,r_{0}=10^{2}\,{\rm pc}$, respectively. On the other hand, the subsonic expansion is not achievable for the case of 
$r_{0}=10\,{\rm pc}$  because the ambient gas density is too low. 

\begin{figure}
\includegraphics[width=0.95\textwidth]{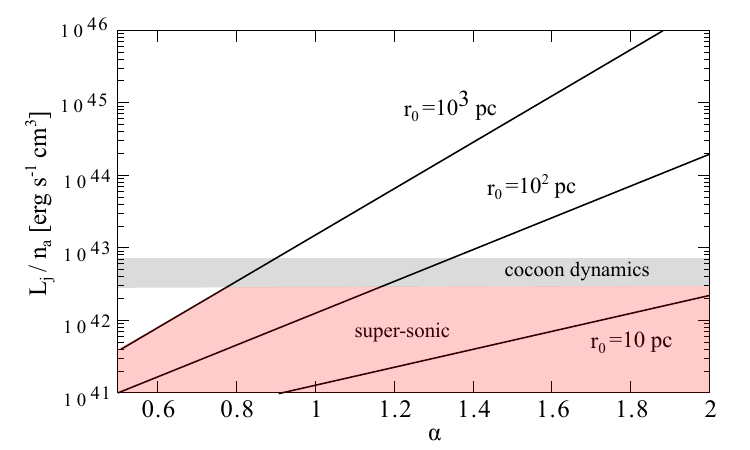}
\caption
{
Same as Figure \ref{fig:5}, but for the condition of the super-sonic expansion of the inner cocoon with $t_{\rm j}=12\,{\rm yr}$ and $R_{\rm h}=1.4\,{\rm pc}$.
}
\label{fig:6}
\end{figure}

Figure \ref{fig:6} shows the supersonic conditions for the inner cocoon.
The black lines show the transonic condition (Eq. (9)) for different critical radii $r_{0}=10\,{\rm pc}$, $100\,{\rm pc}$ and $10^{3}\,{\rm pc}$ 
with $t_{\rm j}=t_{\rm age}=12\,{\rm yr}$ and $R_{\rm h}=1.4\,{\rm pc}$. 
We should note that $t_{\rm j}=12\,{\rm yr}$ is the lower limit because the inner cocoon still expands supersonically at present.
This indicates that the slope index of the ambient density ($\alpha$) must be relatively flat to maintain the supersonic expansion, that is, $\alpha=0.5-0.8$ for $r_{0}=10^{3}\,{\rm pc}$ and $\alpha=0.5-1.2$ for $r_{0}=10^{2}\,{\rm pc}$. 
For $r_{0}=10\,{\rm pc}$, the supersonic condition matches all ranges of $\alpha$, that is, $0.5 < \alpha < 2$. Therefore, to explain the dynamics of both the inner and outer cocoon/shell, 
the critical radius $r_{0}$ should be the range of $10\,{\rm pc} < r_{0} \leq 10^{3}\,{\rm pc}$. 

\begin{figure}
\includegraphics[width=0.95\textwidth]{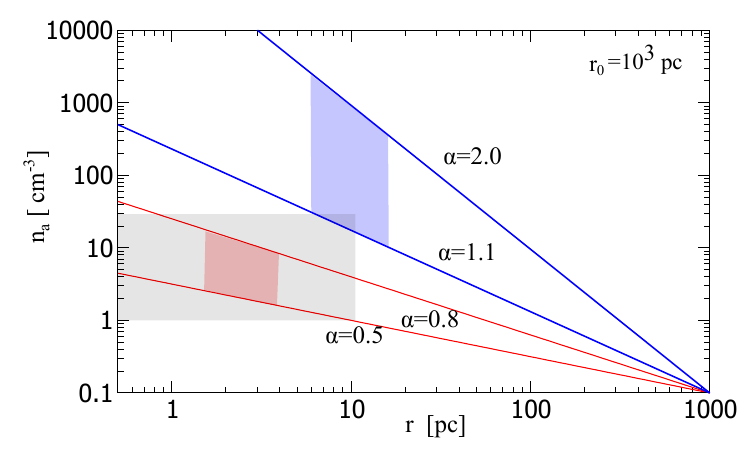}
\caption
{
Predicted diffuse gas density around the inner (red lines) and outer cocoon (blue lines) for $r_{0}=10^{3}\,{\rm pc}$.The red lines corresponds to $\alpha=0.5-0.8$, and the blue lines corresponds to $\alpha=1.1-2.0$, respectively. 
The red and blue shaded regions represent the allowed range of $n_{\rm a}
(R_{\rm h})$ for different size of cocoons, considering the uncertainty of inclination angle $\theta=18^{\circ}-65^{\circ}$. 
The gray shaded area denote the diffuse gas density ($r < 10\,{\rm pc}$) at 28 radio galaxies \citep[]{Fujita16b}.
}
\label{fig:7}
\end{figure} 

Lastly, we discuss the dependence of $T_{\rm a}$ and $R_{\rm h}$. As seen in Eq. (9),  the subsonic condition decreases with $T_{\rm a}$ due to $(L_{\rm j}/n_{\rm a,obs})_{\rm trans}\propto T_{\rm a}$. 
Thus, if $T_{\rm a}=10^{4}\,{\rm K}$, the subsonic solution for the outer cocoon disappears, which contradicts the radio observations, as seen in Figure \ref{fig:MOJAVE_3epochs}. In addition, we assess how our results depend on $R_{\rm h}$. Adopting the upper value of $R_{\rm h}$, we find that the transonic condition is (2-6) times larger than the fiducial cases (Figures \ref{fig:5} and \ref{fig:6}) because of $(L_{\rm j}/n_{\rm a})_{\rm trans}\propto R_{\rm h}^{3-\alpha}$(see Eq. (9)). Thus, this effect does not alter our main conclusion significantly, although the allowed range of $\alpha$ is slightly changed.  

\subsubsection{Allowed $n_{\rm a}(R_{\rm h})$}
\begin{figure}
\includegraphics[width=0.95\textwidth]{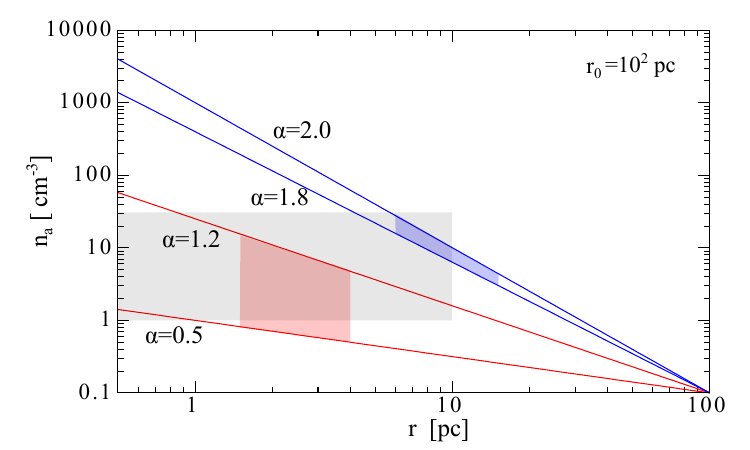}
\caption
{
Same as Figure \ref{fig:6}, but for $r_{0}=10^{2}\,{\rm pc}$.The red lines corresponds to $\alpha=0.5-1.2$, 
and the blue lines corresponds to $\alpha=1.8-2.0$, respectively.
}
\label{fig:8}
\end{figure}

As discussed in Section 3.1.2, we constrain the slope index of ambient gas with $\alpha$ for the outer cocoon (6 pc scale) and the inner cocoon (1 pc scale) by using the transonic condition of cocoon expansion. 
Thus, we here constrain the diffuse gas density around the cocoons as seen in Figures \ref{fig:7} and \ref{fig:8}. The gray-shaded regions represent the allowed range of $n_{\rm a}(R_{\rm h})$ for two different sizes of cocoons, considering the uncertainty in the inclination angle $\theta=18^{\circ}-65^{\circ}$ and in the size of cocoon $R_{\rm h}$. 
Note that the uncertainty in the size of the inner cocoon is $R_{\rm h}=1.4-4\,{\rm pc}$, while that in the size of the outer cocoon is $R_{\rm h}=5.6-16\,{\rm pc}$.
Figure \ref{fig:7} shows that for $r_{0}=10^{3}\,{\rm pc}$, 
the number density is $n_{\rm a}(R_{\rm h})=2-20\,{\rm cm}^{-3}$ at the head of the inner cocoon, while $n_{\rm a}(R_{\rm h})$ is $10-3\times 10^{3}\,{\rm cm}^{-3}$ at the head of the outer cocoon. 
For $r_{0}=10^{2}\,{\rm pc}$, 
we find that the ambient density of the inner cocoon is $n_{\rm a}(R_{\rm h})=0.5-15\,{\rm cm}^{-3}$ and that of $n_{\rm a}(R_{\rm h})=3-30\,{\rm cm}^{-3}$ as seen in Figure \ref{fig:8}. Thus, Figures \ref{fig:7} and \ref{fig:8} indicate that the median gas density around the inner cocoon is at least one order of magnitude lower than that around the outer cocoon. 
Regarding the difference in $n_{\rm a}(R_{\rm h})$ around the inner/outer cocoons, we can interpret that the inner cocoon, recently formed by the jet emitted about 10 years ago, is expanding within the low-density cocoon created by the jet emitted about 50 years ago. 

It is worth to comparing with previous work on the diffuse gas density at central 10 pc region. \cite{Fujita16b} estimated the gas density profiles in the innermost regions of 28 nearby FRI/FRII radio galaxies (including 3C 84) with $n_{\rm a} \simeq 1-30\,{\rm cm}^{-3}$, 
assuming that hot gas outside the Bondi radius is in nearly a hydrostatic equilibrium in a gravitational potential, and the gas temperature near the galaxy center is close to the virial temperature of the galaxy.
Thus, our results are broadly consistent with the range of 
the diffuse gas density $n_{\rm a}(r < 10\,{\rm pc})$ estimated at the center of 28 radio galaxies (including 3C84). On the other hand, for the outer cocoon, we show the possibility of a unusual high number density of the ambient gas with $n_{\rm a} \sim 10^{2}-10^{3}\,{\rm cm}^{-3}$. If this is the case, it would be worth to constraining $n_{\rm a}$ by using other independent method, e.g., the Farady rotation measurement. 

\newpage
\subsection{Estimated $L_{\rm j}$ for inner and outer cocoons}
\begin{figure}
\includegraphics[width=0.95\textwidth]{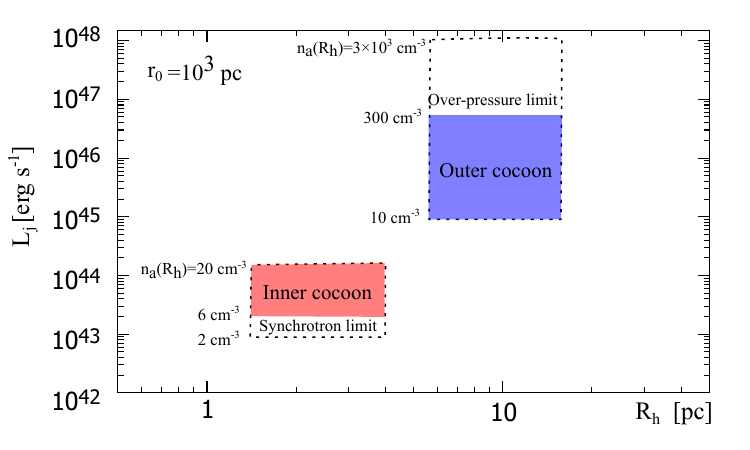}
\caption
{
Total kinetic power is function of the distance from the BH for $r_{0}=10^{3}\,{\rm pc}$. The dotted black square shows the allowed area not including the over-pressure limit and Synchrotron limit. The red region is the average kinetic power of the inner cocoon, while the blue region corresponds to that of outer cocoon. 
The corresponding ambient density $n_{\rm a}(R_{\rm h})$ is also shown at the 
left side of allowed regions (red and blue).
}
\label{fig:9}
\end{figure}

\begin{figure}
\includegraphics[width=0.95\textwidth]{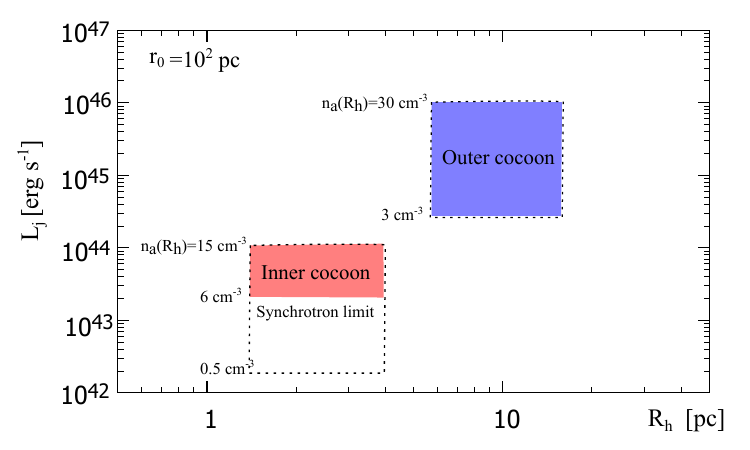}
\caption
{
Same as Figure \ref{fig:8}, but for $r_{0}=10^{2}\,{\rm pc}$. 
}
\label{fig:10}
\end{figure}

By combining $L_{\rm j}/n_{\rm a}(R_{\rm h})$ and $n_{\rm a}(R_{\rm h})$, we can finally evaluate the total kinetic power of the jets ($L_{\rm j}$) for the inner and outer cocoons. Comparing the jet power for the inner and outer cocoons with the different scales, we discuss the time evolution of $L_{\rm j}=L_{\rm j}(t)$. 
Figure \ref{fig:9} shows the total kinetic power of the inner and outer cocoons for $r_{0}=10^{3}\,{\rm pc}$. 
As a result, the total kinetic power of the inner cocoon is $(0.9-14)\times 10^{43}\,{\rm erg}\,{\rm s}^{-1}$, while that of the outer cocoon is $(0.09-5.1)\times 10^{46}\,{\rm erg}\,{\rm s}^{-1}$. We also examine the dependence of $r_{0}$ on $L_{\rm j}$. For $r_{0}=10^{2}\,{\rm pc}$, it is found that $L_{\rm j}$ for the inner cocoon is $(2-11)\times 10^{43}\,{\rm erg}\,{\rm s}^{-1}$ and $L_{\rm j}$ is $(0.03-1.1)\times 10^{46}\,{\rm erg}\,{\rm s}^{-1}$ for the outer cocoon as seen in Figure \ref{fig:10}. 

As seen in Figures \ref{fig:9} and \ref{fig:10}, we constrain their maximum/minimum jet kinetic power as follows. For the minimum jet power, we adopt the observed Synchrotron emission from the inner cocoon. Taking the minimum energy conditions as the base, \citet{Savolainen23} estimated the lower limit of the inner jet power $L_{\rm j, min}=2\times 10^{43}\,{\rm erg}\,{\rm s}^{-1}$. Based on this lower jet power, we can constrain the minimum ambient density around the inner cocoon with $n_{\rm a}(R_{\rm h}=1.4\,{\rm pc})\simeq 6\,{\rm cm}^{-3}$. On the other hand, we constrain the maximum jet power by the condition in which the inner cocoon is over-pressured. 
When the inner over-pressured cocoon can evolve in the cavity formed by the outer jets, the pressure of the inner cocoon $P_{\rm c, inner}$ must be greater than that of the outer cocoon $P_{\rm c, outer}$. From this condition, the maximum ratio of $L_{\rm j,outer}/L_{\rm j, inner}$ is $\simeq 370$. \footnote{
The cocoon pressure $P_{\rm c}$ is obtained as the equation of energy conservation in the cocoon; 
\begin{eqnarray}\label{eq:pc}
\frac{{\hat \gamma}}{{\hat \gamma}-1}
\frac{P_{\rm c}V_{\rm c}}{t_{\rm {age}}}= L_{\rm j}, \nonumber  
\end{eqnarray}
where $P_{\rm c}$ and $V_{\rm c}$ are the total one-side cocoon pressure and the volume of one-side cocoon, respectively. 
Here we assume $\hat \gamma =4/3$, since the cocoon is expected to be dominated by relativistic particles for FRIIs and 3C84 \citep[e.g.,][]{Begelman89,kino07}. 
The volume of the cocoon is measured by $V_{\rm c}=(4\pi/3)R_{\rm h}^{3}$. 
Thus, the maximal ratio of $L_{\rm j,outer}/L_{\rm j, inner}$ 
is given by 
\begin{eqnarray}
\frac{L_{\rm j, outer}}{L_{\rm j, inner}}
=\left(\frac{t_{\rm age, inner}}{t_{\rm age, outer}}\right) 
\left(\frac{R_{\rm h, outer}}{R_{\rm h, inner}}\right)^{3}
\simeq 370, \nonumber 
\end{eqnarray}
where $t_{\rm age, inner}=12\,{\rm yr}$, $t_{\rm age, outer}=55\,{\rm yr}$, $R_{\rm h, inner}=1.4\,{\rm pc}$ and $R_{\rm h, outer}=16\,{\rm pc}$. 
}
Thus, we find that the upper limit of the outer jet power is $L_{\rm j, max}\simeq 5\times 10^{46}\,{\rm erg}\,{\rm s}^{-1}$.
As a result, the jet power of the outer cocoon is $10^{45-46.5}\,{\rm erg}\,{\rm s}^{-1}$, while  that of the inner cocoon is $10^{43-44}\,{\rm erg}\,{\rm s}^{-1}$. 

Overall, we find that the total kinetic power of the older outer cocoon is one or two orders of magnitude higher than that of the inner cocoon. That is, the total kinetic power $L_{\rm j}$ is not constant in time, at least for approximately 50 yr. On the other hand, the observed peak radio flux density of the outer cocoon is $60\,{\rm Jy}$ at 37 GHz \cite[e.g.,][]{Nesterov95}, while that of the inner cocoon is $10\,{\rm Jy}$ at 43 GHz \cite[e.g.,][]{Suzuki12,Nagai12}. In relation to this, the broad H$\beta$ emission line is correlated with the millimeter-wave luminosity of the outer cocoon \citep[][]{Punsly18}. This might indicate that both jet power and accretion power were decreasing. These trends could be roughly consistent with what we found in this paper. 

\begin{figure}
\includegraphics[width=0.95\textwidth]{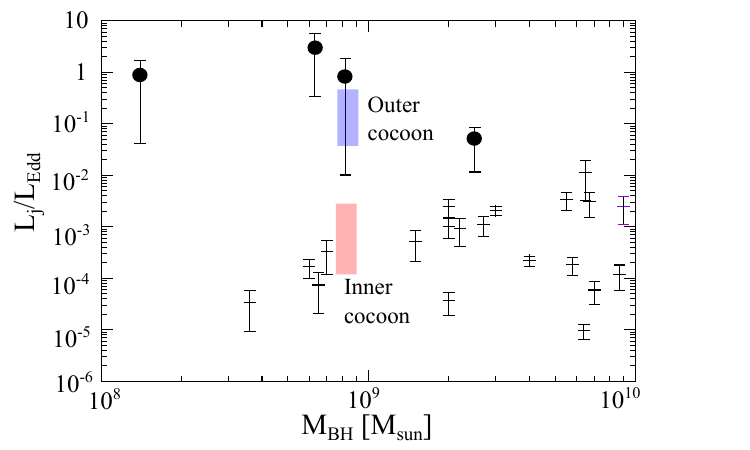}
\caption
{
Jet efficiency $L_{\rm j}/L_{\rm Edd}$ against the BH mass $M_{\rm BH}$. 
The red region is the allowed kinetic energy of the inner cocoon and the blue region corresponds to that of outer cocoon. The black minuses denote the FRI radio galaxies \citep[]{Fujita16b}, and the black circles show the powerful FRII radio galaxies \citep[]{Ito08}. 
}
\label{fig:11}
\end{figure}

In general, the jet power is related to not only the mass accretion rate onto a SMBH but also the magnetic fields around a SMBH and the BH spin\citep[e.g.,][]{Zam14}. However, for the jetted AGNs (radio-loud AGNs), the jet power is proportional to the accretion luminosity \citep[e.g.,][]{Rawlings91, Ghisellini14}. In addition, the general relativistic magnetohydrodynamics simulations found that the poloidal magnetic flux ($\Phi_{\rm BH}\equiv \phi_{\rm BH}\sqrt{\dot{M}_{\rm acc}R_{\rm s}^{2}c}$) in the accretion flow is advected inward until the ram pressure of accretion gas is balanced with the magnetic pressure, where $\phi$ is the dimensionless magnetic flux \citep[e.g., ][]{Tchekhovskoy11,YN14}. 
If the Blandford and Znajeck (BZ) process \citep[]{BZ77}, where the electromagnetic energy extraction from a rotating BH ($\dot{M}_{\rm acc}$) is operated as a jet launching mechanism, the jet power is proportional to the mass accretion onto a SMBH because of $\Phi_{\rm BH}\propto \dot{M}_{\rm BH}^{0.5}$. Thus, postulating that the relativistic jet is mainly powered by the release of gravitational energy ($L_{\rm acc}$) of accreting matter, although the jet power is also related to the BH spin, the jet power ($L_{j}$) can be proportion to $L_{\rm acc}$. Thus,$L_{\rm j}/L_{\rm Edd}$ gives the minimum mass accretion rate normalized by the Eddington mass accretion rate. We here briefly discuss on the efficiency of the jet power ($L_{\rm j}/L_{\rm Edd}$) for the inner and outer cocoons in 3C 84. We found that {\bf $0.01\%-0.2\%$} of the Eddington luminosity $L_{\rm Edd}$ is carried away as the kinetic power of the inner cocoon, where $L_{\rm Edd}=9.6\times 10^{46}\,{\rm erg}\,{\rm s}^{-1}$ with $M_{\rm BH}=8\times 10^{8}M_{\odot}$ \citep[]{Scharwachter13}. On the other hand, $L_{\rm j}$ of the outer cocoon is $3\%-40\%$ of $L_{\rm Edd}$. 
We here compare with $L_{\rm j}/L_{\rm Edd}$ of other radio galaxies as shown in Figure \ref{fig:11}. It is found that 
$L_{\rm j}/L_{\rm Edd}$ of the inner cocoon is similar to that of less powerful FRI radio galaxies \citep[]{Fujita16b} 
and that of the outer cocoon is comparable with that  of powerful FRII radio galaxies \citep[]{Ito08}.
Thus, whether the radio sources are short-lived would be independent of $M_{\rm BH}$ and $L_{\rm j}/L_{\rm Edd}$.
This suggests that the mass accretion may have changed drastically in the last 50 years. In the next section, 
we will discuss how the accretion can change and the origin of the short-lived outer cocoon with high $L_{\rm j}$. 

\section{Discussions}\label{sec:discussions}
Our analysis in Section 3 indicates that the production of a high-luminosity jet ($L_{\rm j}$) over a short period was probably responsible for the generation of the outer cocoon. To achieve this through a change in the accretion rate, a significant and rapid increase in the mass accretion rate would be required, occurring within a brief timescale of a few tens of years to activate the jet.
In Section 4.1, we investigate whether the instability of the accretion disk could naturally account for the required $L_{\rm j}$ within the required timescale of 25-50 years. 
In Section 4.2, we also consider another  possible scenario of tidal disruption events (TDEs) of massive stars as a major source of mass supply \citep[e.g.,][]{Readhead94,Ricci23,Readhead24}. We will discuss these two possibilities as follows.  

\subsection{Scenario of outer cocoon formation driven by disk instability}
One possibly relevant mechanism is thermal instability, associated with changes in the surface density of the disk driven by changes in opacity \citep[e.g.,][]{Lin86, Czerny09}. The state transitions expected for these disk instabilities happen at an Eddington ratio of a few percent, which is observed in some AGNs and black hole binaries \citep[e.g.,][]{Readhead94,An12,Ricci23,Readhead24}. 
To check if the disk instability is valid for the outer cocoon in 3C 84, we examine relevant key physical quantities, i.e., dynamical timescale ($t_{\rm dyn}$), viscous timescale ($t_{\rm vis}$), thermal timescale($t_{\rm th}$), accretion power ($L_{\rm acc}$) and jet power ($L_{\rm j}$).
First, we discuss these timescales for the outer cocoon, assuming the standard accretion disk. 
In general, the variability in time may be caused by the motion of the gas with circular velocity $v_{\rm c}$ within the accretion disk, whose size is $R_{\rm AD}$. 
The dynamical timescale is determined by 
\begin{eqnarray}
t_{\rm dyn}=\frac{R_{\rm AD}}{v_{\rm c}}
\simeq 0.3\,{\rm yr} 
\left(\frac{R_{\rm AD}}{150\,R_{\rm s}} \right)^{3/2}
\left(\frac{M_{\rm BH}}{10^{9}M_{\odot}}\right), 
\end{eqnarray}
where $R_{\rm s}$ is the Schwarzschild radius, $R_{\rm s}=2GM_{\rm BH}/c^{2}$. We here adopt $R_{\rm AD}=150\,R_{\rm s}$ as a fiducial value for the UV-emitting disk size, i.e., the typical outer radius of the accretion disk, assuming a standard thin disk model \citep[e.g.,][]{Kato08}. 
The timescale associated with the disk heating up or cooling down, i.e. the thermal timescale, can be estimated as
\begin{eqnarray}
t_{\rm th}\simeq \frac{t_{\rm dyn}}{\alpha}
\simeq 10\,{\rm yr} 
\left(\frac{\alpha_{\rm vis}}{0.03} \right)^{-1}
\left(\frac{R_{\rm AD}}{150\,R_{\rm s}} \right)^{3/2}
\left(\frac{M_{\rm BH}}{10^{9}M_{\odot}}\right).
\end{eqnarray}
On the other hand, by analogy to black hole X-ray binaries, the variability may be driven by the high-to-soft state transitions \citep[e.g., ][]{Noda18,Ruan19}. The timescale of these transitions can be comparable to the viscous timescale 
\citep[]{Czerny09}
\begin{eqnarray}
t_{\rm vis}\simeq 5\times 10^{3}\,{\rm yr}
\left(\frac{H_{\rm AD}/R_{\rm AD}}{0.05} \right)^{-2}
\left(\frac{\alpha_{\rm vis}}{0.03} \right)^{-1}
\left(\frac{R_{\rm AD}}{150\,R_{\rm s}} \right)^{3/2}
\left(\frac{M_{\rm BH}}{10^{9}M_{\odot}}\right), 
\end{eqnarray}
where $H_{\rm AD}$ and $\alpha_{\rm vis}$ are the scale height and the viscous parameter of the accretion disk \citep[]{Sakura73,Kato08}. 
We use $\alpha_{\rm vis}=0.03$ as a fiducial value because the numerical simulations derive the estimation of $\alpha_{\rm vis}\sim 0.03$ 
\citep[e.g.,][]{Machida00, Machida03, Hirose09,Davis10}. 
For a standard thin disk, the disk aspect ratio ($H_{\rm AD}/r_{\rm AD}$) is typically very small, and we assume $H_{\rm AD}/R_{\rm AD}=0.05$ by following \citet{Stern18}. 
Thus, it is  found that $t_{\rm age}$ is longer than $t_{\rm dyn}$ and shorter than $t_{\rm vis}$ and roughly comparable to $t_{\rm age}$. 
Note that increasing the aspect ratio, $H_{\rm AD}/R_{\rm AD}$ leads to shorter viscous timescales (Eq.(14))\citep[e.g., ][]{Ricci23}. Therefore, in terms of the timescales, the age of the outer cocoon may be explained by these disk instabilities.
Moreover, we should mention that the timescale of fast radio variations occurred in 1959, i.e., a few years, is similar to $t_{\rm dyn}$. This might indicate 
that the mass accretion rate increases drastically caused by the disk instability.

Next we compare the jet power with the accretion power. 
As we already mentioned in \$ 3.2, we here suppose that the accretion disk has a radiative luminosity of $L_{\rm acc}=\epsilon \dot{M}_{\rm acc}c^{2}$, where $\dot{M}_{\rm acc}$ and $\epsilon$ are the mass accretion rate and the radiative efficiency $\epsilon \sim 0.05 -0.4$\citep[e.g.,][]{Sakura73,Balbus91}. 
On the other hand, the jet power is expressed as $L_{\rm jet}=\eta \dot{M}_{\rm acc}c^{2}$ where $\eta=0.1-3$. Especially, $\eta$ reaches maximally $\eta_{\rm max}\sim 3$ for the magnetically arrested accretion (MAD) phase with the BH spin \citep[e.g.,][]{Tchekhovskoy11,narayan22}. 
Thus, a near-maximally rotating BH put roughly 10 times as much power into jets as could be released by the accreated in a steady radiatively efficient flow, i.e., 
\begin{eqnarray}
L_{\rm j}=\left(\frac{\eta}{\epsilon}\right)L_{\rm acc}\simeq 10 L_{\rm acc}, 
\end{eqnarray}
where $\eta\sim 3$ and $\epsilon\sim 0.4$. This may be a proper upper limit, since the jet power in blazer is maximally 10 times greater than the luminosity of their accretion disks\citep[e.g.,][]{Ghisellini14}. 
Thus, the expected maximum jet power $L_{\rm j}$, 
which is derived from the observed mass accretion disk luminosity in 3C 84 with $L_{\rm acc}=4\times 10^{44}\,{\rm erg}\, {\rm s}^{-1}$\citep[][]{Levinson95}, 
is obtained as 
\begin{eqnarray}
 L_{\rm j, max}\simeq 4\times 10^{45}\,{\rm erg}\, {\rm s}^{-1}. 
\end{eqnarray}
This value is one order of magnitude lower than the upper value of jet power with $L_{\rm j}\simeq (1-5)\times 10^{46}\,{\rm erg}\, {\rm s}^{-1}$ (see Figures \ref{fig:9} and \ref{fig:10}). 
This may suggest the extremely high accretion rate, which is one order magnitude higher than the mass accretion rate in the persistent state 
($\dot{M}_{\rm acc}\simeq 6\times 10^{-2}M_{\odot}\,{\rm yr}^{-1}$), 
happened about $25-50$ years ago driven by the disk instability, in order to explain the formation of outer cocoon, where $\dot{M}_{\rm acc}=L_{\rm acc}/(\epsilon c^{2})$ with $\epsilon=0.1$.

\begin{figure}
\includegraphics[width=0.95\textwidth]{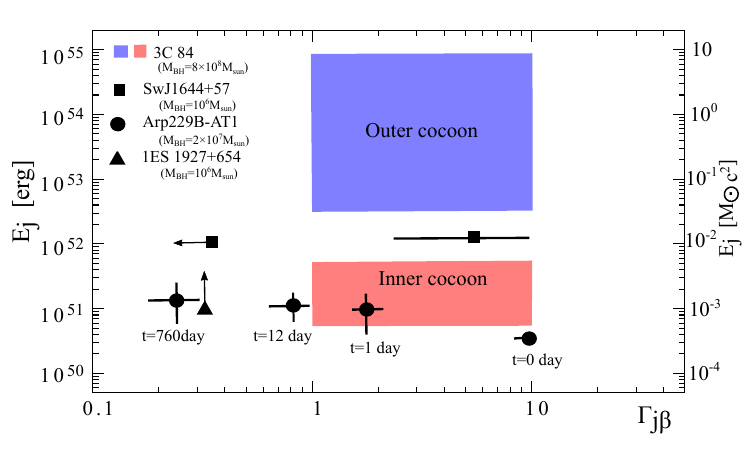}
\caption
{
Jet kinetic energy $E_{\rm j}$ versus the expansion speed
$\Gamma_{\rm j}\beta$, where $\Gamma_{\rm j}=(1-\beta^{2})^{-1/2}$ is the bulk Lorentz factor and $\beta=v/c$. The red and red regions denote the case of 3C 84 jets ($M_{\rm BH}=8\times 10^{8}M_{\odot}$).
The red region is the allowed kinetic energy of the inner cocoon and the blue region corresponds to that of outer cocoon with $\Gamma_{\rm j}\beta =1-10$ \citep[]{Giovannini18}.The right vertical axis denotes the jet kinetic energy in unit of $M_{\odot}c^{2}$. The black square show SwJ1644+57 ($M_{\rm BH}\simeq 10^{6}M_{\odot}$) \citep[]{Zauderer11}, the black circle denotes Arp298-B AT1  ($M_{\rm BH}\simeq 2\times 10^{7}M_{\odot}$) for four different epochs, i.e., form right to left, just after the jet is lunched, one day, 12 days and 760 days \citep[]{Mattila18}, and the black traiangle is 
the upper limit of $E_{\rm j}$ for 1ES 1927 + 654 ($M_{\rm BH}\simeq 10^{6}M_{\odot}$), assuming $L_{\rm j}=10^{43}\,{\rm erg}\,{\rm s}^{-1}$, $t_{\rm age}=3\,{\rm yr}$ and $\beta=0.3$ \citep[]{Meyer24}. 
The arrow is the upper limit for the expansion speed of SwJ1644+57 \citep[]{Yang16}.
}
\label{fig:12}
\end{figure}

\subsection{Scenario of outer cocoon formation driven by tidal disruption events of stars}
In this sub-section, we propose a novel scenario of TDE of a star as a source of mass supply within the short period for the outer cocoon 
of 3C84. 
Some AGN jets are indeed to be suggested to be powered by TDEs, whose light curves are characterized by a power-law decay with an index of $-5/3$. So far, several candidate jets/outflows from TDEs have been observed. For two Swift sources (SwJ1644+57, SwJ2058+05), sudden switch-on and switch-off over a period of two years indicated that jet launching could be associated with super-Eddington accretion \citep[][]{Bloom11,Burrows11}.

Recently, \citet{Mattila18} discovered the spatially resolved relativistic jet structure associated with a bright infrared flare such as Arp299B-AT1 in the nucleus of the Arp 299 galaxy. Recently, multi-wavelength observations support the interpretation of AT2022cmc as a jetted TDE containing a synchrotron afterglow, likely launched by a SMBH with spin \citep[]{Andreoni22}. 
Moreover, the destruction and recreation of the X-ray corona could occur by the TDE like a changing look AGN 1ES 1927 + 654 \citep[]{Ricci20}. Interestingly, this object has recently shown a newly launched radio jet a few years after the destruction of the X-ray corona \citep[]{Meyer24}. 
Theoretically, this may be explained by the rapid increase in the mass accretion rate due to the shocks between the accretion disk and the stellar debris \citep[e.g., ][]{Chan19}.

Statistically, the birth rate of jetted TDEs with $0.02-10^{3}\,{\rm Gpc}^{-3}\,{\rm yr}^{-1}$ \citep[e.g.,][]{Burrows11, DeColle20,Andreoni22} is much higher than the birth rate of compact symmetric objects (CSOs) with $\sim 3\times 10^{-5}\,{\rm Gpc}^{-3}\,{\rm yr}^{-1}$ \citep[][]{Readhead24}, although the intrinsic fraction of jet-accompanying events is largely unconstrained in the range of $0.001\%-34\%$ \citep[]{Kawamuro16}. 
Considering the evidence that most CSOs are short-lived AGNs \citep[]{Kiehlmann24b, Readhead24}, some CSOs would be linked to TDEs. 

Here we discuss the possibility that the properties of the outer cocoon in 3C 84 are explained by TDEs in terms of the timescale and the total jet energy.

Generally, TDEs occur when stars pass within the tidal radius $R_{\rm T}$ at the pericenter of their orbit around an SMBH where $R_{\rm T}$ is determined by the balance between the self-gravity of stars and the tidal force of the SMBH, i.e., $GM_{*}/R_{*}^{3}\simeq (GM_{\rm BH}/R_{\rm T}^{2})\times (R_{*}/R_{\rm T})$. 
The ratio of $R_{\rm T}$ to $R_{\rm s}$ is
\begin{eqnarray}
\frac{R_{\rm T}}{R_{\rm s}}
\simeq 1.6 \left(\frac{M_{*}}{3M_{\odot}}\right)^{-1/3}
\left(\frac{R_{*}}{10R_{\odot}}\right)
\left(\frac{M_{\rm BH}}{10^{9}M_{\odot}}\right)^{-2/3},  
\end{eqnarray}
where $M_{*}$ and $R_{*}$ are the mass and radius of each star undergoing tidal disruption \citep[e.g., ][]{MacLeod12}.
Thus, only massive stars ($M_{*} > 3M_{\odot}$) and/or post-main sequence stars can be observably disrupted by SMBHs with $M_{\rm BH} \sim 10^{9}M_{\odot}$ in 3C 84 \citep[e.g.,][]{Readhead24}. Recently, \citet{Lim20} reported thousands of compact and massive blue star clusters have formed at a steady rate over past $1\,{\rm Gyr}$ around 3C 84. This might support that some massive stars in these star clusters is a major source of mass supply. 

During the encounter, material is stripped from the stellar core and spread into two tidal tails. The material in one of the tails is unbound from the black hole and ejected on hyperbolic trajectories. The other is bound to the black hole and returns on a wide range of elliptical orbits.
Following \cite{Rees88}, we can evaluate the fallback time of the most bound material $t_{\rm fb}$, assuming that the star is initially on a parabolic orbit. 
The characteristic timescale on which post-TDE bound material falls back onto the SMBH is the fallback timescale, $t_{\rm fb}$ as
\begin{eqnarray}
t_{\rm fb}
\simeq 40\,{\rm yr} 
\left(\frac{M_{*}}{3M_{\odot}}\right)^{-1}
\left(\frac{R_{*}}{10R_{\odot}}\right)^{3/2}
\left(\frac{M_{\rm BH}}{10^{9}M_{\odot}}\right)^{1/2}.
\end{eqnarray}
Thus, the TDE timescale is $t_{\rm TDE}\simeq t_{\rm fb}\sim 40\,{\rm yr}$, which is roughly comparable to the age of the short-lived outer cocoon $t_{\rm age}=50\,{\rm yr}$. 
On the other hand, $t_{\rm fb}$ is much longer than the duration of the rapid flare phenomena in the optical band, i.e., several months \citep[]{Nesterov95}. Thus, such the short timescale variability may not be related to TDEs phenomena \citep[e.g.,][]{Andreoni22} but the AGN intrinsic phenomena such as disk instability. 

Next, the maximal accretion rate can be estimated as the ratio $M_{*}/t_{\rm fb}$, 
\begin{eqnarray}
\dot{M}_{\rm max}=\frac{M_{*}}{t_{\rm fb}}
\simeq 0.1M_{\odot}\,{\rm yr}^{-1}
\left(\frac{M_{*}}{3M_{\odot}}\right)^{2}
\left(\frac{R_{*}}{10R_{\odot}}\right)^{-3/2}
\left(\frac{M_{\rm BH}}{10^{9}M_{\odot}}\right)^{-1/2}. 
\end{eqnarray}
The expected maximum jet power associated with TDE, $L_{\rm j,max}=\eta\dot{M}_{\rm max}c^{2}$ is given by
\begin{eqnarray}
L_{\rm j,max}
\simeq 2\times 10^{46}\,{\rm erg}\,{\rm s}^{-1} 
\left(\frac{M_{*}}{3M_{\odot}}\right)^{2}
\left(\frac{R_{*}}{10R_{\odot}}\right)^{-3/2}
\left(\frac{M_{\rm BH}}{10^{9}M_{\odot}}\right)^{-1/2},
\end{eqnarray}
where we adopt $\eta=3$ as an upper value for the MAD phase with the BH spin. This value is comparable to the maximal jet power of the outer cocoon. 
Moreover, the maximum energy $E_{\rm j}$ deposited within the outer cocoon during $t_{\rm age}$  is 
\begin{eqnarray}
E_{\rm j}\simeq 
1.5\times 10^{54}\,{\rm erg} 
\left(\frac{L_{\rm j}}{10^{46}\,{\rm erg/s}}\right)
\left(\frac{t_{\rm age}}{50\,{\rm yr}}\right). 
\end{eqnarray}
Taking into account the uncertainty of $E_{\rm j}$, 
Figure \ref{fig:12} shows that the allowed range of $E_{\rm j}$ corresponds to $3\times 10^{-2}-10M_{\odot}c^{2}$. 
Thus, the total energy embedded in the outer cocoon may be explained by a single massive star capture by a SMBH. 
As seen in Figure \ref{fig:12}, we found that the kinetic energy of the outer cocoon is relatively higher than that of three jetted TDEs such as "SwJ 1644+57"($M_{\rm BH}\simeq 10^{6}M_{\odot}$) \citep[]{Zauderer11}, "Arp299B-AT1" ($M_{\rm BH}\simeq 2\times 10^{7}M_{\odot}$) \citep[]{Mattila18} and 1ES1927 +654 ($M_{\rm BH}\simeq 10^{6}M_{\odot}$) \citep[]{Meyer24} which show a prominent extended jet structure and a rapid decrease in radio luminosity over a few years. The SMBH mass of 3C 84 is two-three orders of magnitude larger, and thus the short-lived jet in 3C84 may be a scale-up version of jetted TDEs because of $E_{\rm j}\propto M_{\rm BH}$. 
Table 2 summarizes the comparison of the properties of the outer cocoon with two possible scenarios. From these, the formation of outer cocoon associated with the extreme accretion events may be driven by the tidal disruption events (TDEs) of massive stars and/or the disk instability.

\begin{deluxetable*}{cccccc}[ht]
\tablecaption{Comparison of the properties of the outer cocoon with two scenarios   \label{table:2}}
\tabletypesize{\small}
\tablehead{
  \colhead{ } &
  \colhead{Outer cocoon} &
  \colhead{Disk instability}
&    \colhead{TDE} 
  }
\startdata
Timescale of variability &  $t_{\rm age}\sim 50\,{\rm yr}$ 
& $t_{\rm dyn}\sim 0.3\,{\rm yr}$, $t_{\rm th}\sim 10\,{\rm yr}$
$t_{\rm vis}\sim 10^{4}\,{\rm yr}$ & 
$t_{\rm fb}\sim 40\,{\rm yr}$ \\
Maximum jet power
$(L_{\rm j, max})$& 
$(1-5)\times 10^{46}\,{\rm erg}\,{\rm s}^{-1}$ 
& $\sim 5\times 10^{45}\,{\rm erg}\,{\rm s}^{-1}$ 
& $\sim 2\times 10^{46}\,{\rm erg}\,{\rm s}^{-1}$  \\
Total deposited
energy ($E_{\rm j}$) & 
$\sim 1.5\times 10^{54}\,{\rm erg}$  
& $-$ 
& $\sim 10^{54}\,{\rm erg}(\simeq M_{\odot}c^{2})$ \\
\enddata
\end{deluxetable*}

\begin{figure}
\includegraphics[width=0.95\textwidth]{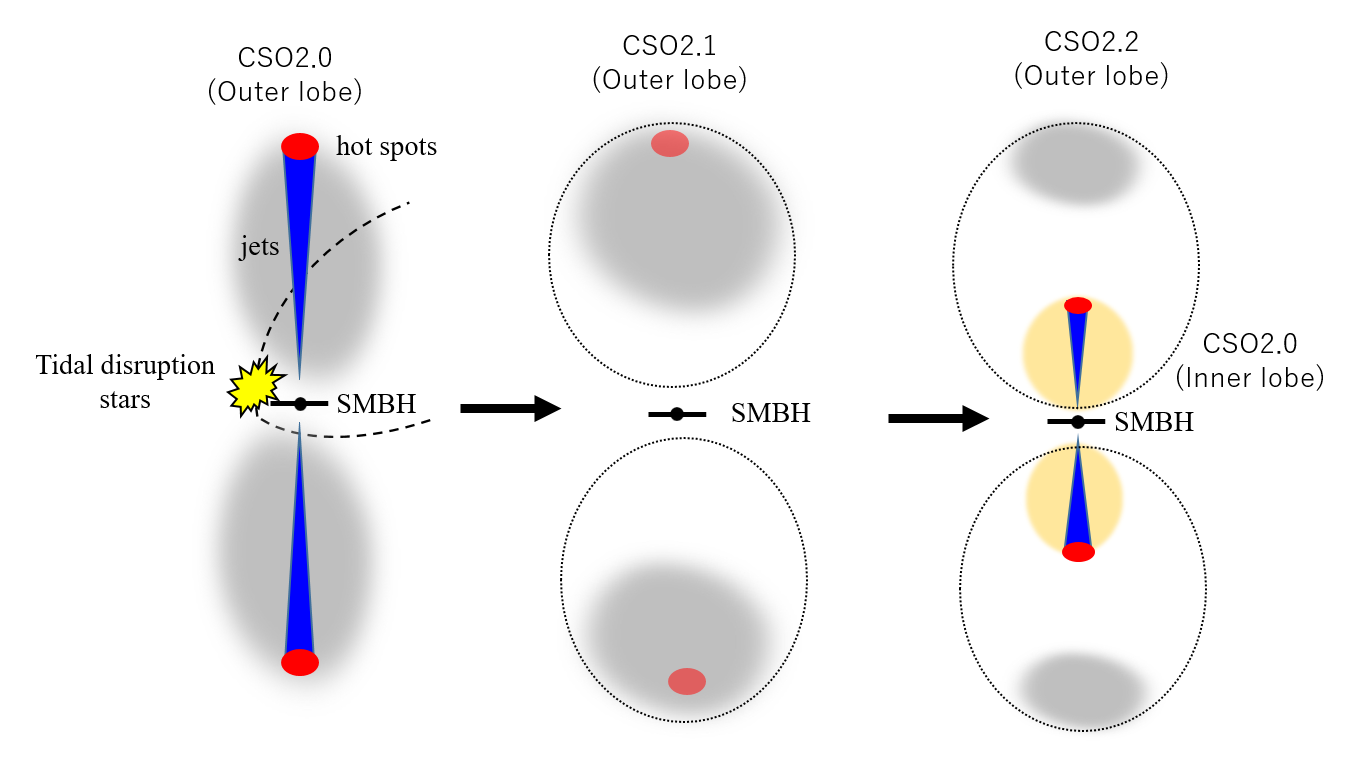}
\caption
{
A diagram of the evolutionary sequence of inner ($\sim 1\,{\rm pc}$) 
and outer ($6\,{\rm pc}$) cocoon in 3C 84 based on the recently proposed evolution model of CSO2 powered by the TDEs \citep[][]{Sullivan24}. 
In the early-life of outer cocoon (CSO 2.0s) , their edge-brightened lobes (gray-colored bubbles) with prominent hot spots (red-colored circles) are formed by a relativistic jets (blue-colored triangles) driven by the interaction between the tidal disruption stars and accretion disk. As the bound gas mass is exhausted (or the accretion disk is destroyed by the interaction) , the jet activity for the outer cocoon must be terminated. After that the outer cocoon transits from the super-sonic expansion to sub-sonic one. This phase corresponds to the mid-life of CSO 2s, namely CSO 2.1. In the late-life of outer cocoon (CSO 2.2s), their expansion is likely subsonic because the apparent motion is invisible and the radio morphology is highly patchy (see also Figure \ref{fig:MOJAVE_3epochs}).
During the late-life of outer cocoon, the new born jets propagate within it, and thus the inner cocoon (small orange-colored bubbles) would be identified as the new CSO 2.0.
}
\label{fig:13}
\end{figure}

Based on the TDE scenario, we next discuss the evolutionary sequence of the inner and outer cocoons in 3C 84 as seen in Figure \ref{fig:13}. 
phase (a): In the early life of the outer cocoon, their edge-brightened lobes (gray-colored bubbles) with prominent hot spots (red-colored circles) are formed by relativistic jets (blue-colored triangles) driven by the stellar TDEs. 
phase (b): After the jet activity for the outer cocoon is terminated, the outer cocoon must transition from relativistic to subsonic.  
phase (c): In the late life of the outer cocoon, their expansion is likely subsonic because the apparent motion is not visible and the radio morphology is highly patchy (see also Figure \ref{fig:MOJAVE_3epochs}). During the late-life of the outer cocoon, the newborn jets propagate within the region of the old, outer cocoon whose density is lower than the ambient gas density around the outer cocoon. 
This idea is supported by analytical cocoon dynamics and simulations \citep[e.g.,][]{Kaiser00, kino07}.

Obserbationally, an edge-brightened high luminosity class CSO named CSO 2 is roughly distributed into three subclasses: CSO 2.0s, CSO 2.1s, and CSO 2.2s. CSO 2.0s have prominent hot spots at the leading edges of narrow lobes, while CSO 2.2s have no hot spots and wider lobes. The size and radio luminosity of CSO 2.0s are smaller and larger than CSO 2.2s. In addition, the expansion velocity of CSO 2.0s is $\sim 0.3$ c (super-sonic), while that of CSO 2.2s is almost negligibly small (sub-sonic). CSO 2.1s represents a hydrid feature between CSO 2.0s and CSO 2.2s. Based on this evidence, an evolutionary sequence CSO 2.0s (phase(a)) $\to$ CSO 2.1s (phase(b)) $\to$ CSO 2.2s (phase(c)) has been proposed \citep[][]{Readhead24, Sullivan24}. 
Considering the total energy of the outer cocoon ($\sim 10^{54}\,{\rm erg}$) and the three epochs of the actual 3C 84 images (the fading radio lobes in Figure \ref{fig:MOJAVE_3epochs}), the physical properties of the outer cocoon resemble the evolutionary sequence of short-lived CSO 2s. Based on the above definition of CSO subclasses, the inner cocoon can also be identified as the new CSO 2.0. Unlike the outer cocoon, the inner cocoon would be in a persistent phase because its jet power $L_{\rm j}=10^{43-44}\,{\rm erg}\,{\rm s}^{-1}$ is comparable to the average jet power in the past ($\sim 10^{7}\,{\rm yr}$) estimated by 10 kpc-scale X-ray cavity, e.g. $L_{\rm j}\simeq 1.5\times 10^{44}\, {\rm erg}\, {\rm s}^{-1}$ \citep[]{Rafferty06}. 

Recently, for TXS 0128 + 554 that are members of CSOs, \citet{Lister20} discovered the lack of compact, inverted spectrum hot spots and an emission gap between the bright inner jet and the outer radio lobe structure. This might indicate that the jets have undergone episodic activity and were relaunched a decade ago, which is similar to the properties of the inner and outer cocoons in 3C 84. Therefore, future VLBI monitoring observations for the inner cocoon in 3C 84 and TXS 0128+554 would be one of the crucial methods to explore the origin and formation of CSO 2.0. 

\section{Summary}\label{sec:summary}
We investigated the jet power and the density of ambient matter in the radio jet 3C 84 by using the momentum balance along the jet axis and the transonic condition for the mini-cocoons observed at two different scales (approximately 1 and 6 parsec scales). Our findings are as follows. 

\begin{enumerate}
\item 
We precisely determined the ratio of jet power to ambient density, $L_{\rm j}/n_{\rm a}$, to be $(0.3-0.7) \times 10^{43}\,{\rm erg}\,{\rm s}^{-1}\,{\rm cm}^3$ for the inner cocoon. Similarly, for the outer cocoon, we found that this value is more than an order of magnitude larger at $(0.9-3.7) \times 10^{44}\,{\rm erg}\,{\rm s}^{-1},{\rm cm}^3$. This indicates that the outer cocoon is formed by a powerful jet that propagates through an ambient gas density of $n_{\rm a}=20-300\,{\rm cm}^{-3}$, with a jet power of $10^{45-46.5}\,{\rm erg}\,{\rm s}^{-1}$. On the other hand, the inner cocoon is formed by a weaker jet with a power of $10^{43-44}\,{\rm erg}\,{\rm s}^{-1}$, propagating through a relatively low-density environment of $n_{\rm a}=6-20\,{\rm cm}^{-3}$.
Regarding the difference in ambient gas density $n_{\rm a}$, it appears to support the hypothesis that the inner cocoon, recently formed about 10 years ago, is expanding within the low-density cocoon created by the jet emitted about 50 years ago. 

\item 
 We found that the jet efficiency ($L_{\rm j}/L_{\rm Edd}$) rapidly decreases from $3-40\%$ (outer cocoon) to $0.01-0.2\%$ (inner cocoon) over $\simeq 50 \,{\rm yr}$. To explain the short-lived outer cocoon with high $L_{\rm j}$, a significant mass accretion, i.e., $\ge 10\%$ of the Eddington mass accretion rate, must be required over a short period ($\sim 50$ yr) to activate the jet. 
 This may suggest the extremely high accretion rate driven by disk instability, which is one order magnitude higher than the averaged mass accretion rate, happened about $25-50$ years ago. 
Alternatively, the maximum energy $E_{\rm j}$ embedded within the outer cocoon during $t_{\rm age}\sim 50\,{\rm yr}$ corresponds to $3\times 10^{-2}-10M_{\odot}c^{2}$. Thus, the total energy embedded in the outer cocoon may be explained by massive stars and/or post main sequence stars captured by a SMBH. 
  
\item 
Following CSO evolutionary scenario proposed by \citet{Sullivan24},
we discuss the evolutionary sequence of the inner and outer cocoons in 3C 84 as follows: (i) In the early life of the outer cocoon (CSO 2.0s), their edge-brightened lobes with prominent hot spots are formed by relativistic jets driven by TDE of stars. (ii) After the jet activity for the outer cocoon is terminated in about 50 years, 
the outer cocoon must transition from relativistic to subsonic. This phase corresponds to the mid-life of CSO 2s, namely CSO 2.1. (iii) In the late life of the outer cocoon (CSO 2.2s), their expansion is likely subsonic because the apparent motion is visible and the radio morphology is highly patchy. During the late life of the outer cocoon, the newly born inner jets can propagate within the region of the old, outer cocoon whose density is lower than the ambient gas density around the outer cocoon. Thus, the inner cocoon could be identified as the new CSO 2.0.

\end{enumerate}

\acknowledgments
We thank the referee for his/her careful read of this manuscript and valuable comments. The authors thank K. Ohsuga, H. Nagai S. Yoshioka for their valuable comments to the discussions. 
This research used data from the MOJAVE database that is maintained by the MOJAVE team \citep[]{Lister18}.
This study was supported by JSPS KAKENHI grant Nos. 19K03918 (NK), JP22H00157 (MK), and 21H04496 (KW). 

\bibliography{bibtex_TDE}{}
\bibliographystyle{aasjourna}

\end{document}